\documentclass[aps,prc,preprint,superscriptaddress,showpacs,amsmath,amssymb,letter]{revtex4-1}
\usepackage{graphicx}
\usepackage{bm}
\usepackage{color}
\newcommand{\pss}[1]{#1}
\newcommand{\referee}[1]{#1}
\newcommand{\refereetwo}[1]{#1}
\begin{document}

\title{Transport coefficients of nuclear matter in neutron star cores}

\author{P.~S.~Shternin}\email{pshternin@gmail.com}
\affiliation{Ioffe Physical-Technical Institute,
Politekhnicheskaya 26, 194021 St.-Petersburg, Russia}
\affiliation{St.-Petersburg State Polytechnical University,
Politekhnicheskaya 29, St.-Petersburg 195251, Russia}
\author{M.~Baldo}
\affiliation{Instituto Nazionale di Fisica Nucleare, Sez. di
Catania, Via S. Sofia 64, 95123 Catania, Italy}
\author{P.~Haensel}
\affiliation{N. Copernicus Astronomical Center, Polish Academy of
Sciences, Bartycka 18, PL-00-716 Warszawa, Poland}

\date{\today}

\begin{abstract}
We calculate thermal conductivity and shear viscosity of nucleons
in dense nuclear matter of neutron star cores in the
non-relativistic Brueckner-Hartree-Fock framework. Nucleon-nucleon
interaction is described by the Argonne v18 potential with
addition of the Urbana IX effective three-body forces. We find
that this three body force leads to decrease of the kinetic
coefficients with respect to the two-body case. The results of
calculations are compared with electron and muon transport
coefficients as well as with the results of other authors.
\end{abstract}


\maketitle

\section{Introduction}
\label{S:intro}
Kinetic coefficients of neutron star cores are important
ingredients in modeling of various processes in neutron stars. As
the most dense stars in the Universe (with densities exceeding
nuclear density $\rho_0\approx 2.8\times 10^{14}$~g/cm$^3$)
neutron stars are widely considered as unique laboratories for
studying properties of super-dense matter under the extreme
conditions unavailable in terrestrial laboratories. Due to this
reason studies of neutron stars attract constant interest.

It is believed that a neutron star consists of the dense core
filled with uniform asymmetric nuclear matter surrounded by the
thin crust (for example, \cite{hpyBOOK}). The outer part of the
core contains neutrons with small admixture of protons, and
electrons and muons as charge-neutralizing component. The equation
of state and composition of inner parts of neutron stars are
poorly known. The different possibilities, apart from the nuclear
matter, are: hyperon core, kaon or pion condensates, quark core.
It is possible that all or some  neutron stars  are in fact
so-called strange stars containing self-bound strange quark
matter.

In what follows we restrict ourselves to the simplest model of the
nucleon neutron star core which consists of neutrons (n), protons
(p), electrons (e), and muons ($\mu$). The nuclear matter is in
the equilibrium state with respect to the beta-processes, which is
commonly called beta-stable nuclear matter.

In the present paper we consider shear viscosity and thermal
conductivity of the neutron star cores. The thermal conductivity
is needed for modeling thermal structure and evolution of such
stars. It is especially important for \pss{studying} cooling of
young neutron stars (age $\lesssim 100$~yr) where the internal
thermal relaxation is not yet finished (e.g.,
\cite{gyp01,sy08b_eng}). Shear viscosity is important for studying
decay of the oscillations of neutron stars and stability of
rotating stars \pss{(e.g. \cite{ak01})}.

\pss{The} diffusive kinetic coefficients are governed by the
particle collisions. \pss{The} first detailed \pss{studies} of the
kinetic coefficients in neutron star cores were made by Flowers
and Itoh \cite{fi79}. They considered n-p-e matter taking into
account collisions between all particle species. Flowers and Itoh
constructed the exact solution of the multicomponent system of
transport equations. The small amount of protons and small
magnitude of \pss{the} electron-neutron interaction lead to the
conclusion that the kinetic coefficients can be split in two
almost independent parts -- the neutron kinetic coefficients
mediated by nucleon-nucleon collisions and electron kinetic
coefficients mediated by the collisions between charged particles;
the proton kinetic coefficients are small.  The up-to-date
electron and muon contribution to kinetic coefficients of neutron
star cores \pss{was} calculated by Shternin and Yakovlev
\cite{sy07} (thermal conductivity) and \cite{sy08b} (shear
viscosity). Here we will focus on  the neutron kinetic
coefficients.

Flowers and Itoh \cite{fi79} based their calculations on the
\pss{free} nucleon scattering amplitudes, which were derived from
the experimentally determined phase shifts. They neglected the
Fermi-liquid effects and nucleon many-body effects. The results of
Flowers and Itoh \cite{fi79} were later reconsidered in
Refs.~\cite{bhy01,sy08b,bh99}. In latter works it was assumed that
\pss{the} main medium effects are incorporated in the effective
masses, while the free-space nucleon potential was used.

However in \pss{a} strongly interacting dense matter the many-body
effects play most important role. These effects in the context of
transport coefficients of pure neutron matter were first addressed
in Refs.~\cite{wap93,sbrs94}. In Ref.~\cite{wap93} the attempt of
\pss{a} consistent many-body consideration of the kinetic
coefficients on the basis of Fermi-liquid theory was made. Authors
of Ref.~\cite{sbrs94}  used the concept of thermodynamical
$T$-matrix, neglecting Fermi-liquid effects.

A decade \pss{later} the medium modifications of the neutron star
matter transport coefficients were reconsidered in
Refs.~\cite{bv07,cb11,zlz10,bpvv10} \pss{on the basis of} modern
realistic nucleon interactions. Different approaches were used. In
Ref.~\cite{bv07} the correlated basis function (CBF) formalism
were incorporated to obtain the shear viscosity of pure neutron
matter. Later the same group \cite{bpvv10} compared the transport
coefficients obtained from CBF and  the Brueckner-Hartree-Fock
(BHF) $G$-matrix formalism in pure neutron matter and found a good
agreement between the results of both approaches;  Carbone and
Benhar \cite{cb11} used CBF formalism to calculate transport
coefficients of beta-stable nuclear matter which is directly
related to the neutron star properties. Zhang et al. \cite{zlz10}
calculated the transport coefficients in the framework of the BHF
theory.

The general result of Refs.~\cite{bv07,cb11,zlz10,bpvv10}  is that
the medium effects strongly  increase the values of kinetic
coefficients.

In the present paper we reconsider the problem of the nucleon
kinetic coefficients in the dense mater \pss{in beta equilibrium}
in the BHF framework. Generally, our approach is similar to that
used by Zhang et al. \cite{zlz10}. The difference within the
approaches will be emphasized below.

The paper is organized as follows. We start from outlining the
formalism for calculating the kinetic coefficients in
multi-component Fermi-liquid (Section~\ref{S:kincoeff}). In
Sec.~\ref{S:Gmat} we discuss \pss{the} adopted model of the
nucleon interaction and calculate the in-medium nucleon-nucleon
cross-sections. We discuss the results and compare them with those
by  other authors in Sec.~\ref{S:discuss}. Our conclusions are
presented in  Sec.~\ref{S:conclusion}.

\section{Kinetic coefficients}
\label{S:kincoeff}
Let us shortly describe the expressions needed to obtain the
kinetic coefficients. The transport properties of strongly
interacting matter are customary described in the framework of
Landau Fermi-liquid theory (e.g., \cite{bp91}). Consider a
multicomponent Fermi-liquid which consists of quasiparticles of
different species c with distribution functions
$F_c(\mathbf{p}_c)$, where $\mathbf{p}_c$ is the
\pss{quasiparticle} momentum. In equilibrium quasiparticle
distribution functions are \pss{given by} the Fermi-Dirac function
\begin{equation}
  F_c(\mathbf{p}_c)=f_c(\mathbf{p}_c)=\left[1+\exp\left(\frac{\epsilon_c(\mathbf{p}_c)-\mu_c}{k_BT}\right)\right]^{-1},
\end{equation}
where $\epsilon_c(\mathbf{p}_c)$ is the quasiparticle energy,
$\mu_c$ is its chemical potential, $T$ is \pss{the} temperature,
and $k_B$ is the Boltzmann constant. When perturbations, such as
gradients of temperature or \pss{a} hydrodynamical velocity
\pss{$\mathbf{V}$}, are applied to the system, distribution
functions start to deviate from equilibrium \pss{ones}. It is
convenient to present perturbed distribution function in the form
\begin{equation}\label{eq:Phic}
F_c=f_c-\Phi_c\frac{\partial f_c}{\partial \epsilon_c},
\end{equation}
where functions $\Phi_c$ describe this deviation. These functions
depend on quasiparticle quantum numbers and  on the type of the
perturbation \pss{applied} to the system. In order to find
$\Phi_c$ one solves multicomponent system of linearized kinetic
equations, which has the following form for the problems of
thermal conductivity $\kappa$ and shear viscosity $\eta$
\begin{equation}\label{eq:kineq}
  \left.\begin{array}{ll}
  \kappa: & (\epsilon_1-\mu_1){\bf v}_1 \nabla T/T\\
  \eta: &
  \left(v_{1\alpha}p_{1\beta}-\frac{1}{3}\delta_{\alpha\beta}v_1
  p_1\right) V_{\alpha\beta}
  \end{array}\right\} \frac{\partial f_1}{\partial
  \epsilon_1}=\sum_i I_{ci}(12;1'2'),
\end{equation}
where $v_1$ is the quasiparticle velocity and $V_{\alpha\beta}$ is
the rate of strain tensor.\pss{ The latter tensor is defined as
\cite{ll10eng}
\begin{equation}\label{eq:ratestrain}
V_{\alpha\beta}=\frac{1}{2}\left(\frac{\partial V_\alpha}{\partial
x_\beta}+\frac{\partial V_\beta}{\partial x_\alpha}\right),\qquad
\alpha,\ \beta=x,\,y,\,z,
\end{equation}
 where it is assumed that $\mathrm{div} \mathbf{V} = 0$.} The right-hand
side in Eq.~(\ref{eq:kineq}) contains the sum of the linearized
Boltzmann collision integrals describing collisions of
quasiparticles of species c and i:
\begin{widetext}
\begin{eqnarray}
I_{ci}&=&\frac{1}{(1+\delta_{ci})k_B T}
\sum_{\sigma_{1'}\sigma_{2}\sigma_{2'}} \int\int\int \frac{{\rm
d}\mathbf{p}_{1'} {\rm d}\mathbf{p}_{2}{\rm d}
\mathbf{p}_{2'}}{(2\pi\hbar)^9}
w_{ci}(12;1'2') f_1 f_2 (1-f_{1'})(1-f_{2'})\nonumber\\
&&\times
\left(\Phi_{1'}+\Phi_{2'}-\Phi_1-\Phi_2\right),\label{eq:bolz_coll}
\end{eqnarray}
\end{widetext}
where $w_{ci}(12;1'2')$ is the differential transition rate. Here
by labels 1 and 2 we, as usual, denote quasiparticle states before
collisions, while the labels with primes correspond to the final
states. Once the functions $\Phi_c$ are found, the thermal
conductivity and shear viscosity are obtained from the expressions
for the heat current and the dissipative part of the stress
tensor, respectively. These coefficients can be written as
\begin{equation}\label{eq:kappa}
  \kappa_c=\frac{\pi^2 T n_c \tau^{(\kappa)}_c}{3 m_c^*},
\end{equation}
\begin{equation}\label{eq:eta}
  \eta_c=\frac{n_c p_{\rm Fc}^2\tau_c^{(\eta)}}{5m^*_c},
\end{equation}
where $n_c$ is the number density of particles of the c
\pss{species}, $m^*_c$ is their effective mass at the Fermi
surface, and effective relaxation times $\tau^{(\kappa)}_c$ and
$\tau^{(\eta)}_c$ are introduced, which are determined by $\Phi_c$
functions.

In the limit of \pss{low} temperatures, all quasiparticles can be
placed on the Fermi surface where possible. In addition is it
assumed that the transition probability is independent of the
energy transferred in the collision event. The exact solution of
the kinetic equation for one-component Fermi-liquid in form of
rapidly converging series were constructed by Brooker and Sykes
\cite{bs68,sb70} and H{\o}jg{\aa}rd Jensen et al. \cite{hjsw68}.
For multicomponent system the exact solution were given by Flowers
and Itoh \cite{fi79}. Later this approach was further developed in
Ref.~\cite{apq87}. However in order to study general behavior of
kinetic coefficients it is enough to employ much simpler
variational solution of the system of kinetic equations and
introduce correction factors \pss{needed to obtain} the exact
solution. Mathematically, variational solution corresponds to the
first term in the series expansion of the full solution
\cite{bp91}. Below we show that for nuclear matter the difference
between exact and variational solutions is of the order of 20\%
for thermal conductivity and less than 5\% for shear viscosity.

Now let us present the expressions for simple variational solution
of Eqs.~(\ref{eq:kineq})--(\ref{eq:bolz_coll}) for neutron-proton
matter. Below we closely follow the formalism of
Refs.~\cite{bhy01, sy08b}. In this approximation the effective
relaxation times are obtained from the $2\times 2$ system of
algebraic equations
\begin{equation}\label{eq:var_kin}
  \sum_{i=n,p}\nu_{ci} \tau_i=1,\qquad c=n,p,
\end{equation}
where effective collision frequencies $\nu_{ci}$ are expressed in
terms of some effective cross-sections. For thermal conductivity
one writes \cite{bhy01}
\begin{equation}\label{eq:nu_kappa}
  \nu^{(\kappa)}_{ci}=\frac{64 m_c^* m_i^{*2} (k_B T)^2}{5 m_N^2
  \hbar^3}S_{\kappa ci},
\end{equation}
while for shear viscosity one obtains \cite{sy08b}
\begin{equation}\label{eq:nu_eta}
  \nu^{(\eta)}_{ci}=\frac{16 m_c^* m_i^{*2} (k_B T)^2}{3 m_N^2
  \hbar^3}S_{\eta ci},
\end{equation}
where $m_N$ is the bare nucleon mass.

The quantities $S_{\kappa ci}$, $S_{\eta ci}$ with the dimension
of area are the effective cross-sections given by angular
averaging of transition probability with corresponding angular
weight functions. Due to momentum conservation and the fact that
all quasiparticles are placed on the Fermi surface, the relative
positions of four momenta (two initial and two final) involved in
collision are determined only by two angles. Therefore the
transition probability depends on two angular variables. In the
Fermi-liquid theory, traditionally, so-called
Abrikosov-Khalatnikov angles are used. However, they are not so
convenient when dealing with the collisions of particles of
different kind \cite{apq87}. It is possible to use any two
\pss{variables} which are suitable for a given problem and which
fix the relative positions \pss{of momenta}. Instead of
Abrikosov-Khalatnikov angles we selected transferred momentum $q$
($\mathbf{q}=\mathbf{p}_1'-\mathbf{p}_1$) and total colliding pair
momentum $P$ ($\mathbf{P}=\mathbf{p}_1+\mathbf{p}_2$). It turns
out that this choice is most convenient for the BHF calculations.
We note, that the  variable $q$ is connected with the c.m.
scattering angle $\theta_{\rm c.m.}$ as
\begin{equation}\label{eq:theta_cm}
  \cos \theta_{\rm c.m.}=1-\frac{q^2}{2p^2},
\end{equation}
where $p$ is the absolute value of the colliding pair c.m.
momentum $\mathbf{p}\equiv(\mathbf{p}_2-\mathbf{p}_1)/2$. At the
Fermi surface the latter is connected with $P$ via the relation
$4p^2+P^2=2(p_{\rm Fc}+p_{\rm Fi})$.

Utilizing these variables, \pss{the} effective cross-sections
$S_{ci}$ are
\begin{widetext}
\begin{eqnarray}
S_{\kappa cc}&=& \frac{m_N^2}{128\pi^2 \hbar^4 p_{\rm Fc}^3}
\int_0^{2p_{\rm Fc}} {\rm d }P\int_0^{q_m(P)} {\rm d} q\frac{
(4p_{\rm Fc}^2-P^2)}{\sqrt{q_m^2-q^2}} {\cal Q}_{cc}(P,q),\label{eq:Skappa_cc}\\
S_{\kappa ci}&=& \frac{m_N^2}{128\pi^2 \hbar^4 p_{\rm Fc}^3}
\int_{|p_{\rm Fc}-p_{\rm Fi}|}^{p_{\rm Fc}+p_{\rm Fi}} {\rm d
}P\int_0^{q_m(P)} {\rm d} q\frac{
(4p_{\rm Fc}^2+q^2)}{\sqrt{q_m^2-q^2}} {\cal Q}_{ci}(P,q), \quad c\neq i,\label{eq:Skappa_ci}\\
S_{\eta cc}&=& \frac{3 m_N^2}{128\pi^2 \hbar^4 p_{\rm
Fc}^5}\int_0^{2p_{\rm Fc}} {\rm d }P\int_0^{q_m(P)} {\rm d}
q\frac{
q^2(4p_{\rm Fc}^2-P^2-q^2)}{\sqrt{q_m^2-q^2}} {\cal Q}_{cc}(P,q),\label{eq:Seta_cc}\\
S_{\eta ci}&=& \frac{3 m_N^2}{128\pi^2 \hbar^4 p_{\rm
Fc}^5}\int_{|p_{\rm Fc}-p_{\rm Fi}|}^{p_{\rm Fc}+p_{\rm Fi}} {\rm
d }P\int_{0}^{q_m(P)} {\rm d} q\frac{ q^2(4p_{\rm
Fc}^2-q^2)}{\sqrt{q_m^2-q^2}} {\cal Q}_{ci}(P,q),\quad c\neq
i,\label{eq:Seta_ci}
\end{eqnarray}
\end{widetext}
where
\begin{equation}\label{eq:qm}
  q^2_m(P)=\frac{4p_{\rm Fc}^2 p_{\rm Fi}^2 - (p_{\rm Fc}^2+p_{\rm Fi}^2-P^2)^2}{P^2}
\end{equation}
is the square of the maximum possible momentum which can be
transferred in collision at a given value of $P$. In \pss{the}
case of collisions of identical particles (\pss{when} $p_{\rm
Fc}=p_{\rm Fi}$), the Eq.~(\ref{eq:qm}) reduces to much simpler
relation $q_m=2 p$, and $\theta_{\rm c.m.}$ ranges from $0$ to
$\pi$ for any $P$. In general case, there exists a global maximum
c.m. scattering angle which \pss{is realized} when
$\mathbf{p}\perp\mathbf{P}$. This maximum angle can be found from
the relation $\tan \theta_{\rm c.m.}^{\rm max}/4 = p_{\rm
Fi}/p_{\rm Fc}$.

 Note that here we
slightly changed definition of Ref.~\cite{sy08b}, by doubling
$S_{\eta ci}$ in Eq.~(\ref{eq:Seta_ci}) and correspondingly saving
factor 2 in Eq.~(\ref{eq:nu_eta}).

The quantities ${\cal Q}_{ci}$ in
Eqs.~(\ref{eq:Skappa_cc})--(\ref{eq:Seta_ci}) are the squared
matrix elements of the transition amplitude, summed over spin
variables ${\cal Q}_{ci}=1/4\sum_{spins} \left|\langle 12|T|1'2'
\rangle\right|^2$ \cite{bhy01}, where the momentum conserving
delta-function is already taken out. More precisely, it is
connected with the differential transition rate by the expression
\begin{equation}\label{eq:W_Q}
  \sum_{spins} w_{ci}(12;1'2')\equiv 4\frac{(2\pi)^4}{\hbar} \delta(\epsilon_1+\epsilon_2-\epsilon_{1'}-\epsilon_{2'})\delta(\mathbf{P}-\mathbf{P'}) {\cal
  Q}_{ci}.
\end{equation}
Let us stress at this point that the averaged transition
probability $W$ commonly used in the Fermi-liquid theory
\cite{bp91} is given by $W=\pi {\cal Q}_{ci}$.

If the quantities ${\cal Q}_{ci}$ are known, then the expressions
(\ref{eq:kappa})--(\ref{eq:Seta_ci}) are sufficient for calculating
kinetic coefficients in the variational approximation. We have
also calculated the exact solution by solving numerically the
system of kinetic equations.

\section{Transition probability in the Brueckner theory}
\label{S:Gmat}
Two central ingredients needed for the transport theory are the
squared quasiparticle transition amplitude ${\cal Q}_{ci}$ (or
$w_{ci}(12;1'2')$) and the quasiparticle effective mass $m^*_c$.
In what follows we obtain both these quantities in the framework
of non-relativistic Brueckner-Hartree-Fock approximation.

\subsection{Brueckner-Hartre-Fock approximation}\label{S:BHF}
The Brueckner theory proved itself as one of the successful
approaches for treating many-body effects in constructing equation
of state of nuclear matter. The concept of Brueckner $G$-matrix is
described elsewhere (e.g \cite{baldobook}). Here we outline it
briefly focusing on the explicit expressions used. Within this
approach the infinite series of certain terms
(diagrams) in the perturbation expansion for the total energy of
the system are encapsulated in the so-called $G$-matrix operator
to be used then instead of the bare nucleon interaction in the
remaining terms (with care for the double counting of the same
contributions). The main advantage of the $G$-matrix approach is
that $G$-matrix matrix elements do not diverge as it could happen
when using a bare nucleon potential. In the total angular momentum
(partial wave) representation the $G$-matrix depends on the
nucleon  pair total momentum $P$, spin $S$, and angular momentum
$J$. All these quantum numbers are conserved during
the interaction. In addition, the $G$-matrix depends on the
relative momentum and the orbital momentum of the pair before
($p$,$l$) and after ($p'$, $l'$) interaction, respectively, and is
not diagonal in these variables. Finally, the $G$-matrix depends
on the type of collision $\alpha$ (nn, np, or pp) or,
equivalently, on the total isospin projection onto the z-axis in
the isospin space.

The $G$-matrix is determined from the Brueckner-Bethe-Salpeter
equation, which in the $JlS$ representation reads
\begin{widetext}
\begin{equation}\label{eq:Gmat_l}
    G^{\alpha JS}_{ll'}(P,p,p';z)= V^{\alpha
    JS}_{ll'}(p,p')+\sum_{l''}\int k^2 {\rm d} k V^{\alpha
    JS}_{ll''}(p,k)
   \frac{\overline{Q}(P,k)}{z-\overline{E}(P,k)}G^{\alpha
   JS}_{l''l'}(P,k,p';z),
  \end{equation}
\end{widetext}
where $ V^{\alpha JS}_{ll'}(p,p')$ is the matrix element of the
bare nucleon potential, $\overline{Q}(P,k)$ and
$\overline{E}(P,k)$ are the Pauli operator and energy of the
nucleon pair, respectively, both averaged over the direction of
the total momentum $\mathbf{P}$. It can be shown that the use of
angle-averaged operators here is a good approximation
\cite{shiller99,shiller99err}. The parameter $z$ in
Eq.~(\ref{eq:Gmat_l}) is the so-called starting energy which
originates from the energy denominators of the perturbation
expansion.

The particle spectrum in the Brueckner theory is given by
\begin{equation}\label{eq:sp_spect}
  \epsilon(p)=\frac{p^2}{2m_N}+U(p),
\end{equation}
where $U(p)$ is auxiliary self-consistent potential. Originally
this potential was selected to be zero above the Fermi surface
$p>p_F$ and to be determined self-consistently below the Fermi
surface by the expression
\begin{equation}\label{eq:spp}
  U(p_1)=\sum_{p_2<p_{\rm F2}} \langle 12|G(z=\epsilon_1+\epsilon_2)| 12
  \rangle_{\rm A},
\end{equation}
where $12\rangle_A$ means that the wavefunction is properly
anti-symmetrized. It was shown, however, that the so-called
continuous choice of the single particle potential in which it is
given by Eq.~(\ref{eq:spp}) above the Fermi surface ($p_1>p_{\rm
F1}$) as well, minimizes the contribution from the three-hole
lines diagrams (next terms in the cluster expansion for the total
energy) \cite{sbgl98}. We will adopt the continuous choice of the
$U(p)$ throughout the paper. The final result of the
Brueckner-Hartree-Fock  (BHF) approximation is the expression for
the total energy per nucleon
\begin{equation}\label{eq:Etot}
  E/A=E_{\rm kin}+ \frac{1}{2}\sum_\alpha\sum_{p_1<p_{\rm F1}; p_2<p_{\rm F2}}
  \langle 12| G^\alpha(z=\epsilon_1+\epsilon_2)|12\rangle_{\rm A},
\end{equation}
where $E_{\rm kin}$ is the kinetic energy part. In
Eq.~(\ref{eq:Etot}) the summation is carried over all nucleon
Fermi-seas.

The BHF approximation is generated by the bare two-body
interaction $V_{\rm NN}$. It is well known that there exist
essential three-body nucleon interactions  $V_{\rm NNN}$ (which
can not be reduced to the two-body ones). The three-body
interactions are required to obtain correct binding energies of
the few-body systems as well as the correct position of the
symmetric nuclear matter saturation point. The three-body
interaction is included in the BHF equations by means of an
effective two-body interaction $V_{\rm NN}^{(3)}$ which results
from averaging of $V_{\rm NNN}$ over the third particle.
\refereetwo{ It was argued, that the contribution from the
three-body potential should be introduced in different forms (with
different symmetry factors) into Eqs. (21) and (20)
\cite{hebeler10,carbone13}. However, we follow the method of
Refs.~\cite{blatt74,grange76,coon79,martzolf80,ellis85,grange89,bf99,llsz08},
and assume that the three-body forces arise from the inclusion of
non-nucleonic degrees of freedom. The force is reduced to a
density dependent two-body force by averaging on the nucleonic
line along which such degrees of freedom are excited. The average
is weighted by the probability of the particle to be at a given
distance from the other two particles. This probability is
calculated at the two-body level, and it takes into account the
effect of the antisymmetry and the Pauli blocking, as well as of
the repulsive core, which enforces the probability to be
vanishingly small at short distance. In this way only the two
particles are equivalent and of course they are then
antisymmetrized. The criticism of Refs.~\cite{hebeler10,carbone13}
on the the symmetry factors, needed at Hartree-Fock level, looks
not pertinent for this scheme. Furthermore the procedure avoids
possible double counting in the self-consistent procedure for the
single-particle potential.}

\refereetwo{The question can be raised if one should go beyond the
standard BHF approximation and include higher-order terms in
expansion of the single-particle potential, so-called
rearrangement terms. However, this would require a careful
reconsideration of the EOS. In fact, according to the BBG
expansion, additional contributions to the single particle
potential requires the accurate examination of the higher order
terms beyond BHF, at least of the three hole-line diagrams (se,
e.g., Refs.~\cite{baldo90, baldo07, baldoburgio12, taranto13}, and
references therein). It has to be stressed the present BHF EOS is
compatible with the known phenomenological constraints
\cite{taranto13}.}

In the described context the nuclear matter in BHF approximation
with continuous choice of the single particle potential can be
understood as the Fermi-sea of the ``particles'' placed in the
self-consistent field $U(p)$. These ``particles'' have the
momentum-dependent (and density-dependent) effective mass
\begin{equation}\label{eq:m_eff_bru}
 m^*(p)=\left(\frac{1}{p}\frac{{\rm d} \epsilon(p)}{{\rm d}
 p}\right)^{-1},
\end{equation}
and their scattering is governed by the Brueckner $G$-matrix. In
our calculations we do not distinguish the quasiparticles of the
Fermi-Liquid theory (Sec.~\ref{S:kincoeff}) and these
``particles'' in the vicinity of the Fermi surface.
Correspondingly we will use $m^*(p_F)$ as the quasiparticle
effective mass and the on-shell $G$-matrix at the Fermi surface as
the quasiparticle scattering amplitude.

\subsection{Transition probability}\label{S:transprob}
It is straightforward to derive the expression for the
spin-averaged squared matrix elements ${\cal Q}_{ci}$ in
Eq.~(\ref{eq:W_Q}) from the $G$-matrix matrix elements in the
$JlS$ representation. The quantities ${\cal Q}_{ci}$ depend on two
angular variables (Sec.~\ref{S:kincoeff}), and in this
representation are naturally expanded in the Legendre polynomials
${\cal P}_L \left(\cos \theta_{\rm c.m.}\right)$ of the cosine of
the c.m. scattering angle $\theta_{\rm c.m.}$:
\begin{equation}\label{eq:Q_legendre}
  {\cal Q}_{ci}(q,P)=\sum_{L} {\cal Q}^{(L)}_{ci}(P) {\cal
  P}_L\left(\cos \theta_{\rm c.m.}\right).
\end{equation}
The standard angular momentum algebra leads to the following
expression for the coefficients in the above expansion:
\begin{widetext}
\begin{eqnarray}
  {\cal Q}_{ci}^{(L)}(P)&=& \frac{1}{16 \pi^2}\sum i^{\ell'-\ell+\bar{\ell}-\bar{\ell}'}
  \Pi_{\ell\ell'\bar{\ell}\bar{\ell}'}\Pi^2_{J\bar{J}}
  C^{L'0}_{\ell' 0 \bar{\ell}'0} C^{L0}_{\ell0\bar{\ell}0}
  \left\{\begin{array}{ccc}
    \bar{\ell} & S & \bar{J}\\
    J & L & \ell
  \end{array}\right\}
 \left\{\begin{array}{ccc}
    \bar{\ell}' & S & \bar{J}\\
    J & L & \ell'
  \end{array}\right\}\nonumber\\
  &&\times
  G^{JS}_{\ell\ell'}(P,p,p;z) \left(G^{\bar{J} S}
  _{\bar{\ell}\bar{\ell}'}(P,p,p;z)\right)^*\label{eq:Q_L},
\end{eqnarray}
\end{widetext}
where terms in curly brackets are 6$j$-symbols,
$C^{L0}_{\ell0\bar{\ell}0}$ is the Clebsch-Gordan coefficient,
$\Pi_{ab}\equiv\sqrt{(2a+1)(2b+1)}$, and summation is carried over
all angular momenta and spin variables, except $L$. The $G$-matrix
amplitudes $G^{JS}_{\ell\ell'}(P,p,p;z)$ (\pss{collision type}
index $\alpha=ci$ is omitted for brevity) must be taken on shell
($z=\epsilon(p)$) and on the Fermi surface. This ensures that they
depend only on the $P$ variable. For the collisions of like
particles, additional symmetrization factor
\begin{equation}
\left(1-(-1)^{S+\ell+1}\right)\left(1-(-1)^{S+\bar{\ell}+1}\right)
\end{equation}
should be included in Eq.~(\ref{eq:Q_L}) which accounts for the
interference between indistinguishable scattering channels $12\to
1'2'$ and $12\to 2'1'$.

Once the Brueckner-Bethe-Salpeter equation
(\ref{eq:Gmat_l})--(\ref{eq:sp_spect}) is solved and matrix
elements $G^{JS}_{\ell\ell'}$ are found, the effective
cross-sections $S_{ci}$ and, therefore, kinetic coefficients are
obtained by introducing Eqs.~(\ref{eq:Q_legendre})--(\ref{eq:Q_L})
in Eqs.~(\ref{eq:Skappa_cc})--(\ref{eq:Seta_ci}). Note, that the
integration over $q$ in these expressions can be done analytically
leaving one with a single numerical integration over $P$ (see
Appendix~\ref{A:angular}).

\subsection{In-medium cross-section}\label{S:crossect}
It is common to illustrate the many-body effects on the particle
collisions by calculating the in-medium cross-section. However,
the in-medium cross-section is not very well-defined quantity. The
reason is that the Pauli blocking invalidates the usual form of
the optical theorem for the in-medium scattering matrix. In order
to construct the correct in-medium unitary relations the
generalized density of states which include Pauli blocking should
be used \cite{srs90}. In addition, the cross-section depends on
the \pss{motion} state of the colliding pair with respect to the
medium. To avoid these complications it is customary to use the
effective cross-sections defined in the certain way (see, for
example \cite{thm87}). In the context of the transport theory all
particles are placed on the Fermi surface. It is clear that the
inclusion of the Pauli blocking in the outgoing channel will lead
to zero cross-section at $T=0$ (due to diminishing phase space for
the collision). Therefore by the differential in-medium
cross-section (for the unpolarized scatterers) we call the
quantity
\begin{equation}\label{eq:cross}
  \frac{{\rm d} \sigma_{ci}}{{\rm d}\Omega_{\rm c.m.}} \equiv \frac{m^{*2}_{ci}}{16\pi^2 \hbar^4}
  {\cal Q}_{ci},
\end{equation}
taken at the Fermi-surface. This definition includes the effect of
Pauli blocking in the intermediate states only. In addition, the
reduced effective mass $m^{*}_{ci}$ describes the in-medium phase
space modification. The reduced effective mass is defined as
\begin{equation}\label{eq:meff_red_our}
  m_{ci}^*=\frac{2 m_c^*m_i^*}{m_c^*+m_i^*}.
\end{equation}
More rigorous definition would have come from the dependence of
the total energy of the pair
$\epsilon(p)=\epsilon_c(p_1)+\epsilon_i(p_2)$ on the pair c.m.
momentum $p$ \cite{srs90}
\begin{equation}\label{eq:meff_red_schmidt}
  m^*_{ci}=\left(\frac{1}{2p}\frac{{\rm d} \epsilon}{{\rm d}
  p}\right)^{-1}.
\end{equation}
This definition coincides with  Eq.~(\ref{eq:meff_red_our}) in
case when the momenta of the colliding particles are equal and
differs from it in the opposite case. For simplicity we will
always use Eq.~(\ref{eq:meff_red_our}).

Similarly we define the (effective) total in-medium cross-section
as
\begin{equation}\label{eq:cross_tot}
\sigma_{ci} = \frac{1}{1+\delta_{ci}}\int\limits_{(4\pi)} {\rm d}
\Omega_{\rm c.m.}   \frac{{\rm d} \sigma_{ci}}{{\rm d}\Omega_{\rm
c.m.}},
\end{equation}
 where the factor in front of the integration
takes into account double counting of the final states in case of
like particles.

Note that while the Eqs.~(\ref{eq:Q_legendre})--(\ref{eq:Q_L}) are
formally defined for all values of $\theta_{\rm c.m.}$, there
exists maximum possible c.m. scattering angle leaving particles on
the Fermi surface (see Eq.~(\ref{eq:qm})). All values of the
scattering angle are possible only in particular case of particles
with same Fermi momenta. Therefore the total cross-section given
by Eq.~(\ref{eq:cross_tot}) should be treated with caution.

\section{Results and discussion}
\label{S:discuss}

In our calculations we used the full Argonne v18 two-body
potential \cite{argonnev18} which is designed to accurately
reproduce the experimental nucleon scattering phase shifts.
\pss{When considering} the proton-proton scattering the
electromagnetic part is ignored. The effective three-body
interaction is based on the Urbana \cite{urbana1,urbana2} model.
\refereetwo{The parameters of the Urbana interaction are adjusted
to give the correct value of the symmetric nuclear matter
saturation point. Particularly, we used the Urbana IX version of
this three-body interaction.}

The Brueckner-Bete-Salpeter equation with this input was solved
numerically in the iterative process of obtaining self-consistent
potential until the convergence \pss{was} reached. Below we
present the results of calculations paying separate attention
\pss{to} the effects the inclusion of three-body forces on the
kinetic coefficients.

\subsection{Energy and in-medium cross-sections}
\begin{figure}
\includegraphics[width=0.8\columnwidth]{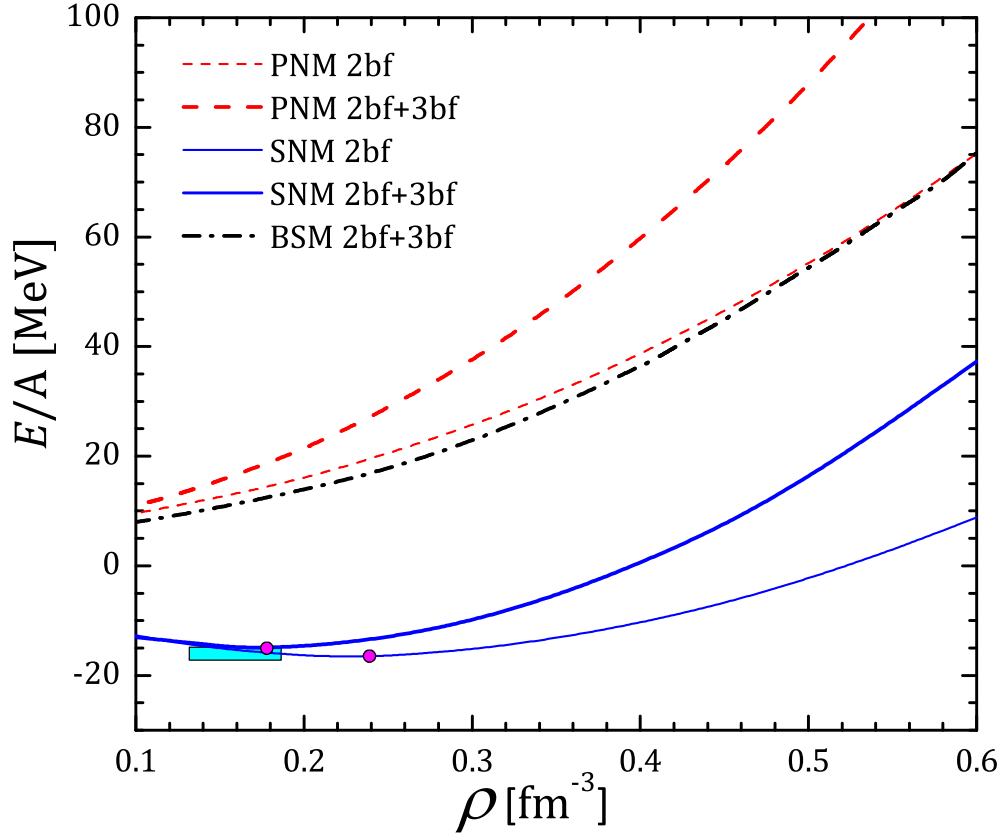}
  \caption{(color online) Energy per nucleon in dense matter as function of baryon density for three states of nuclear matter --
  symmetric nuclear matter (SNM, solid lines), pure neutron matter (PNM, dashed lines), and beta-stable nuclear matter (BSM, dash-dotted lines).
  Thick lines are calculated including two-body
  and three-body nucleon interaction, while thin lines are obtained using two-body forces
  alone. Circles show positions of the saturation points on
  the SNM curves; the experimental position of the saturation point is given by filled rectangle.}\label{F:EOS}
\end{figure}

To begin with, we calculate the total energy per nucleon of
nuclear matter $E/A$ in our model. It is shown in Fig.~\ref{F:EOS}
as a function of density. Energies calculated for symmetric
nuclear matter and pure neutron matter are shown by solid and
dashed lines, respectively. \pss{In} both \pss{cases the} results
calculated with two-body forces only are shown with thin lines in
Fig.~\ref{F:EOS}. Filled area shows the experimental position of
the saturation point of the symmetric nuclear matter. The
calculated positions of the saturation point which are the minima
on the symmetric nuclear matter energy curves are shown by small
circles. Thus the Fig.~\ref{F:EOS} illustrates the well-known
conclusion that two body forces alone do not produce correct
saturation point.

\begin{figure}
\includegraphics[width=0.8\columnwidth]{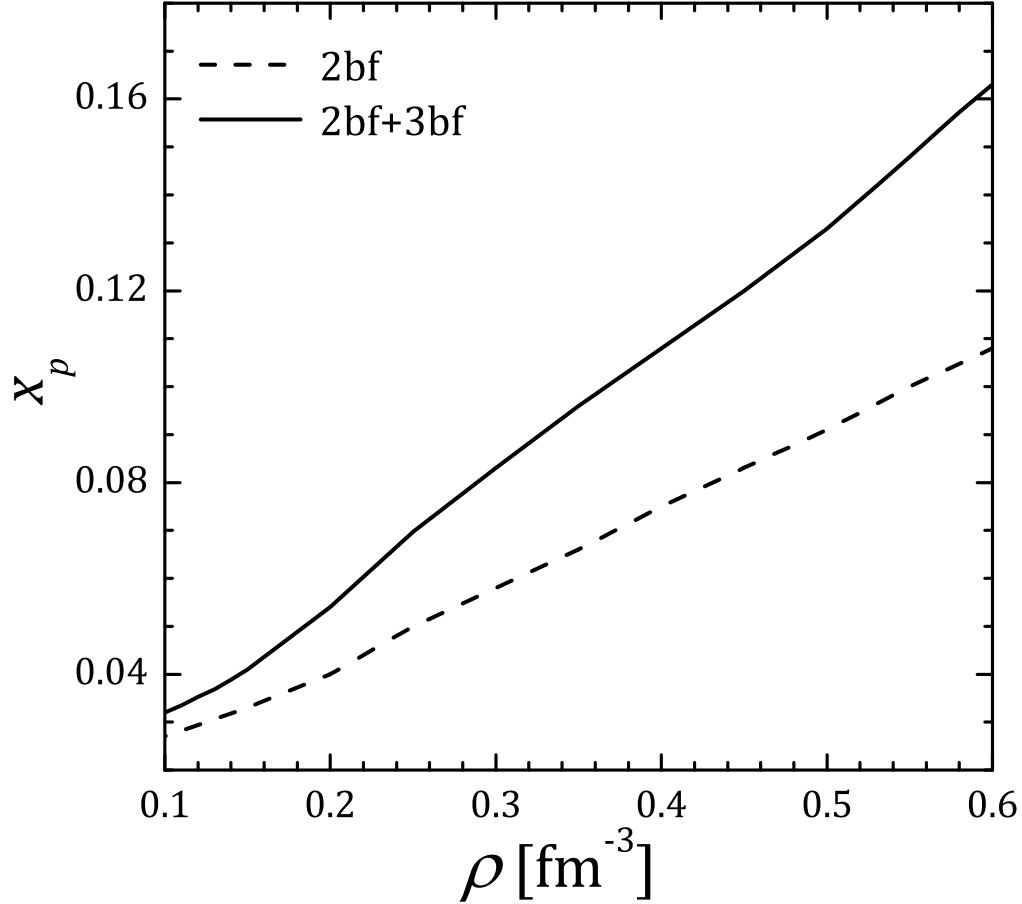}
  \caption{(color online) Proton fraction in beta-stable nuclear matter as function of baryon density.
  Dashed line shows results obtained with two-body forces only.}\label{F:xp}
\end{figure}

The proton fraction in the beta-stable nuclear matter can be
obtained utilizing the quadratic approximation for the energy per
nucleon
\begin{equation}\label{eq:en_quad}
  E=E_0+S_b(1-2 x_p)^2.
\end{equation}
This approximation is known to be very accurate up to $x_p=0$.
Proton fraction resulting from our calculations is shown in
Fig.~\ref{F:xp} with solid line. The dashed line in the same
figure correspond to the proton fraction obtained with two-body
forces only. It is seen that inclusion of three-body forces
increases the proton fraction. In what follows in all calculations
(including those which are referred as ``two-body alone'') we
always use $x_p$ obtained with both two-body and three-body
forces. The energy per nucleon corresponding to this proton
fraction is shown in Fig.~\ref{F:EOS} with dash-dotted curve.

\begin{figure}
\includegraphics[width=0.8\columnwidth]{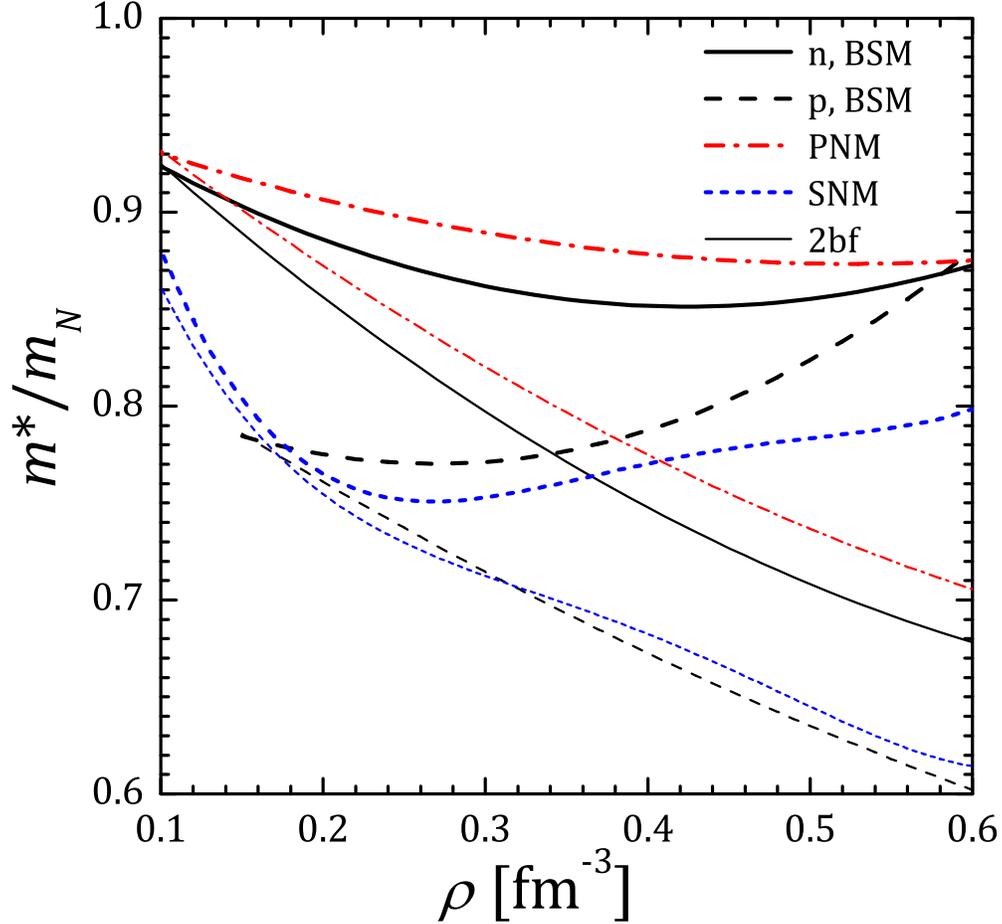}
  \caption{(color online) Effective masses at Fermi surfaces for three states of nuclear matter.
  Thin lines are obtained with two-body forces alone, while thick lines are calculated including two-body and three-body contributions.
  Solid, short dashed, and dot-dashed lines show $m^*_n$ in BSM, SNM, and PNM,
  respectively, while longer dashed lines show $m_p^*$ in BSM.}\label{F:meff}
\end{figure}

The effective masses at the Fermi surface are calculated in
accordance with Eq.~(\ref{eq:m_eff_bru}). It turns out that the
numerical differentiation of the single particle potential
\pss{produces some fluctuations in the dependence} of the
effective masses with density. We thus interpolated the numerical
values by smooth functions of density within 2\% accuracy. The
values obtained in this way are shown in Fig.~\ref{F:meff}. The
neutron effective masses as a function of density for beta-stable,
symmetric, and pure neutron matter are shown with solid, short
dashed, and dot-dashed lines, respectively. Dashed lines show the
proton effective mass in the beta-stable nuclear matter. Thin
lines correspond to the results obtained with two-body interaction
alone. We see, that the medium effects generally decrease the
effective masses from the bare value, but inclusion of the
three-body forces significantly increase the effective masses with
respect to the two-body level. \refereetwo{ Different results were
obtained by Zhang et al.~\cite{zang07,zlz10} who included \pss{an}
additional rearrangement contribution to the effective mass}. The
authors report that this term, resulting mainly from the strong
density-dependence of the effective three-body force $V^{(3)}_{\rm
NN}$, leads to significant decrease of the effective masses. In
what follows we do not include the rearrangement contribution, see
discussion in \refereetwo{Secs.~\ref{S:BHF} and \ref{S:compar}}.

Now we turn to  the in-medium cross-sections calculated in
accordance with Eqs.~(\ref{eq:cross}) and (\ref{eq:cross_tot}).
The cross-sections are parameterized by the quantity $E_{\rm
c.m.}\equiv p^2/m_{N}$ which would be the c.m. energy in the
free-space. As an example we selected one density value
$\rho=0.35$~fm$^{-3}$, approximately  twice the nuclear saturation
density.
\begin{figure*}
\hskip-3cm\includegraphics[width=0.85\textwidth]{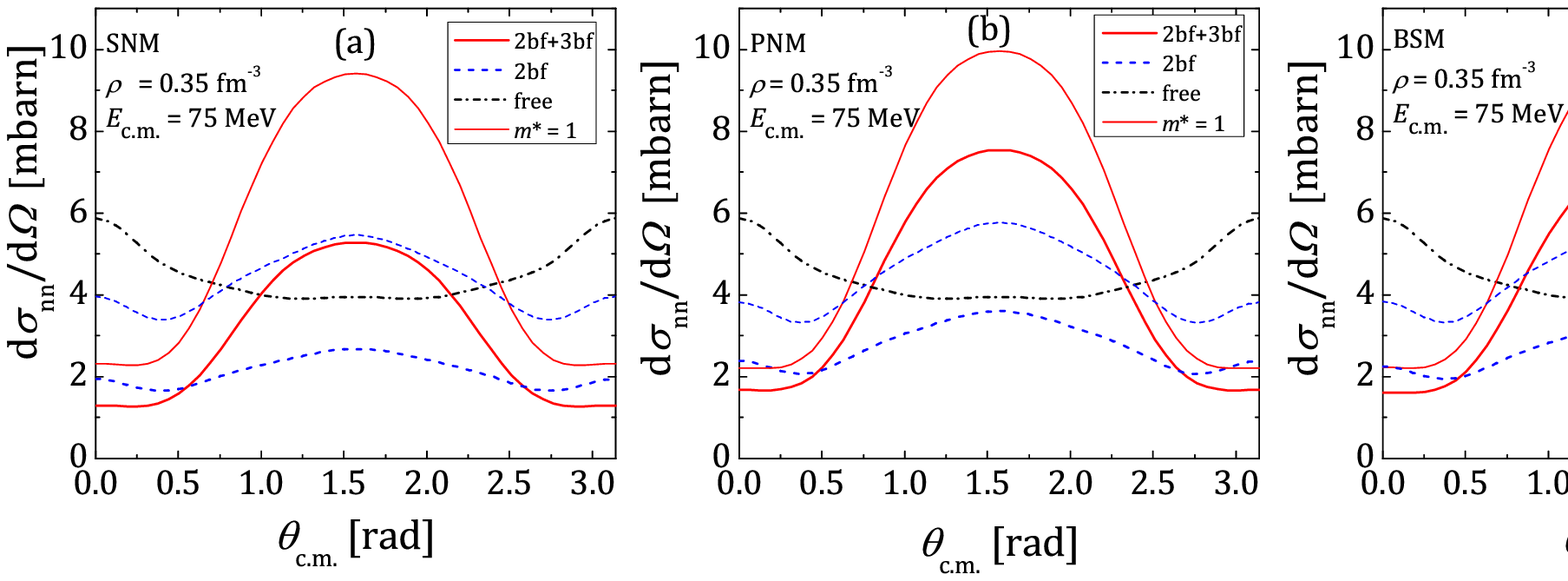}
 \caption{(color online) Differential neutron-neutron cross-section
 as a function of the c.m. angle
  at density $\rho=0.35$~fm$^{-3}$ and c.m. energy $E_{\rm
  c.m.}=75$~MeV in symmetric nuclear matter (panel (a)), pure
  neutron matter (panel (b)), and beta-stable nuclear matter (panel
  (c)). Dash-dotted lines show the free-space cross-sections.
  Solid lines are calculated with two-body and three body
  interactions, dashed lines show the results of calculations with
  two-body potential only. Thin lines correspond to $m^*=1$ case.
  }\label{F:sigmadiffnn}
\end{figure*}
In the Figure~\ref{F:sigmadiffnn} we show neutron-neutron
differential cross-section as a function of the c.m. scattering
angle. Three panels (a), (b), and (c) correspond to three
considered states of the nuclear matter -- symmetric nuclear
matter, pure neutron matter, and beta-stable nuclear matter,
respectively. In each panel the free-space cross-section is shown
by dot-dashed line. Thick dashed lines show the cross-sections
obtained with the two-body potential only. These cross-sections
are smaller than the free-ones. The in-medium suppression is
higher in symmetric nuclear matter than in pure neutron matter and
beta-stable neutron matter. By the thin lines in
Fig.~\ref{F:sigmadiffnn} the cross-sections obtained from
Eq.~(\ref{eq:cross}) are shown but with bare nucleon mass used in
place of the effective mass. Comparison between thin and thick
lines shows that it is the effective mass that is responsible for
the suppression of the cross-section. The situation changes when
three-body forces are included (solid curves in
Fig.~\ref{F:sigmadiffnn}). The inclusion of the three-body forces
increases the cross-sections from the two-body level. It becomes
comparable to and even higher than the free-space cross-section.
The situation is qualitatively similar for the neutron-proton
cross-section. The latter is shown in Fig.~\ref{F:sigmadiffnp} for
symmetric nuclear matter and beta-stable nuclear matter in panels
(a) and (b), respectively. Trivially, there is no np cross-section
in pure neutron matter.

Zhang et al.~\cite{zang07,zlz10}, in contrast, found that
inclusion of three-body forces decreases the in-medium
cross-section from the two-body level. There are two main reasons
for this. First, the three-body force used by the authors of
Refs.~\cite{zang07, zlz10} differs from ours. We have checked that
the inclusion of the three-body force of that type indeed decrease
the cross-section. However, the three-body force model used in
\cite{zang07, zlz10} faces some difficulties in reproducing the
saturation point \cite{llsz08,llszcm06}.  The second reason is in
different approaches to the effective masses. The strong
rearrangement decrease of the effective masses leads to
corresponding decrease of the in-medium cross-sections.

\begin{figure*}
\hskip-2cm\includegraphics[width=0.8\textwidth]{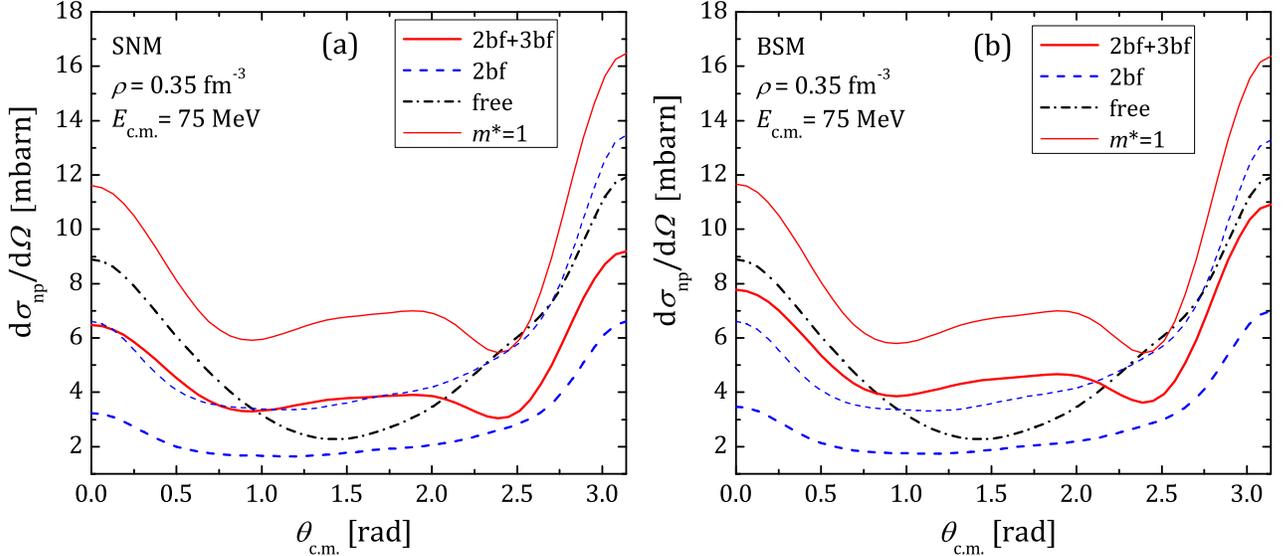}
 \caption{(color online) Differential neutron-proton cross-section
 as a function of the c.m. scattering angle
  at density $\rho=0.35$~fm$^{-3}$ and c.m. energy $E_{\rm
  c.m.}=75$~MeV in symmetric nuclear matter (panel (a)) and beta-stable nuclear matter (panel
  (b)). Notations are same as in Fig.~\ref{F:sigmadiffnn}.
  }\label{F:sigmadiffnp}
\end{figure*}

\begin{figure*}
\hskip-3cm\includegraphics[width=0.8\textwidth]{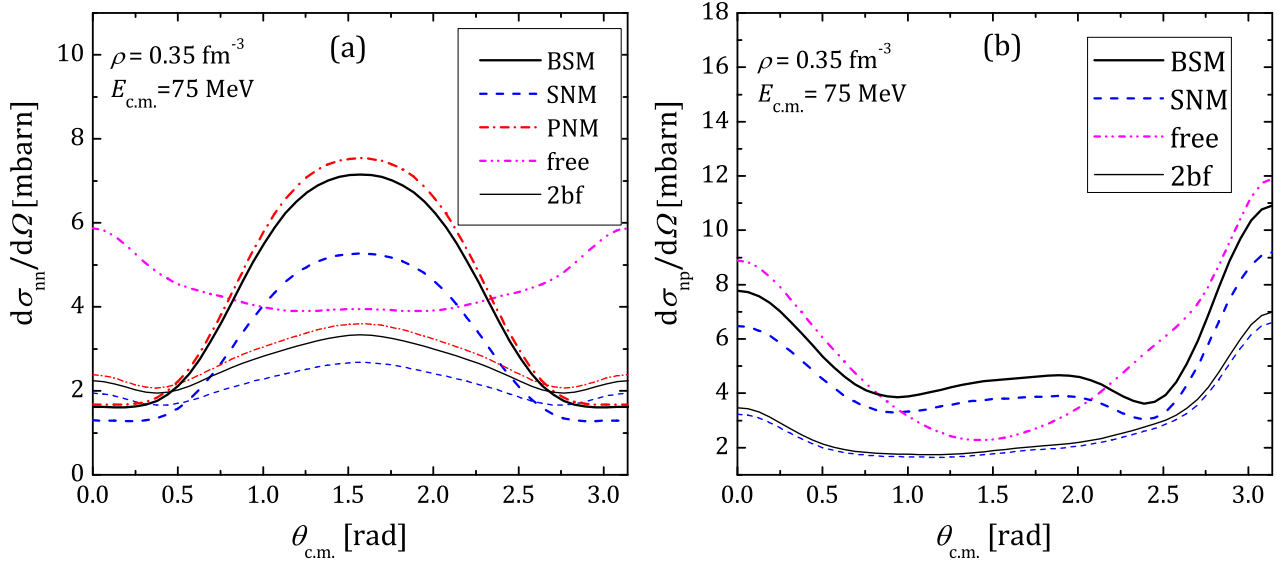}
 \caption{(color online) Differential  neutron-neutron (panel (a)) and neutron-proton (panel (b))
 cross-sections
 as a function of the c.m. scattering angle
  at density $\rho=0.35$~fm$^{-3}$ and c.m. energy $E_{\rm
  c.m.}=75$~MeV in different states. Dot-dot-dashed lines show the free-space cross-section.
  Solid lines correspond to beta-stable nuclear matter, dashed
  lines to symmetric nuclear matter, and dot-dashed lines to pure
  neutron matter. Thin lines show the results obtained with
  two-body potential alone.
  }\label{F:sigmadiffstates}
\end{figure*}

In the Figure~\ref{F:sigmadiffstates} we compare the in-medium
cross-sections in the matter with different nuclear asymmetry.
Neutron-neutron cross-sections are shown in the (a) panel, and
neutron-proton ones in (b) panel. Double dot-dashed lines show
free-space cross-sections. Dot-dashed, dashed, and solid lines are
for PNM, SNM, and BSM, respectively. Thin lines show the two-body
results. We see, that the in-medium nn cross-sections in pure
neutron matter and beta-stable neutron matter are close. The
reason for that is the small proton fraction $x_{p}$. However,
this does not necessary mean that the kinetic coefficients in the
PNM and beta-stable matter \pss{will be close}, see below.

\begin{figure*}
\hskip-3cm\includegraphics[width=0.8\textwidth]{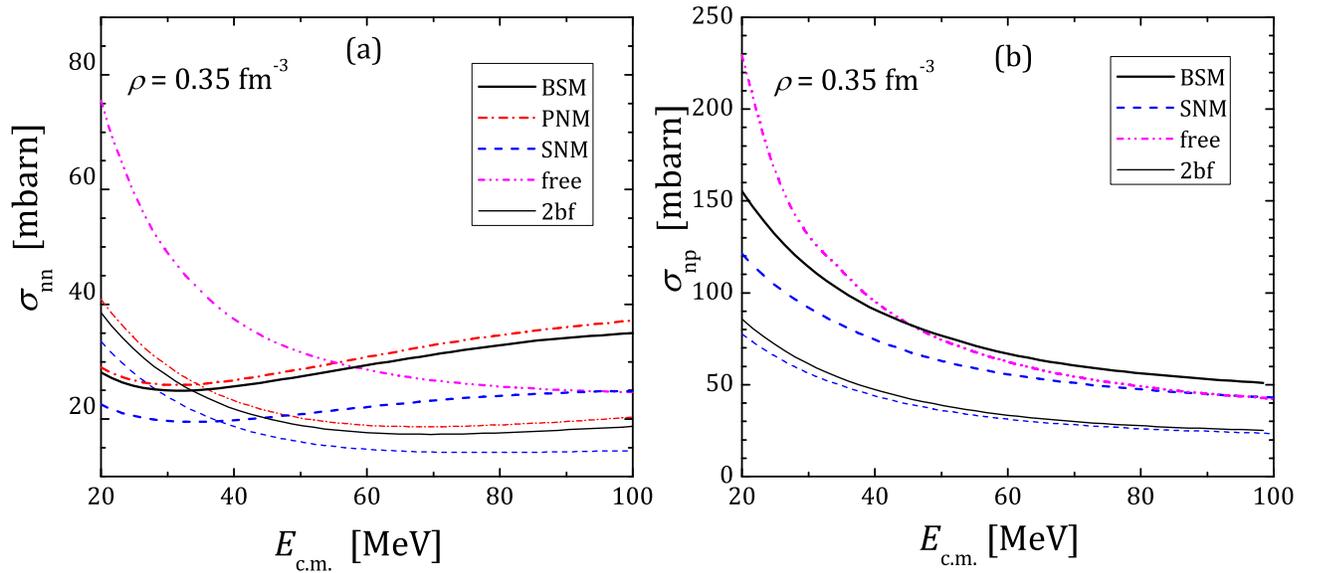}
 \caption{(color online) Total  neutron-neutron (panel (a)) and neutron-proton (panel (b))
 cross-sections  at density $\rho=0.35$~fm$^{-3}$ as a function of c.m. energy $E_{\rm
  c.m.}$ in different states. Notations are the same as in
  Fig.~\ref{F:sigmadiffstates}
  }\label{F:sigmatotstates}
\end{figure*}

Finally in the Fig.~\ref{F:sigmatotstates} we plot the total
neutron-neutron (panel (a)) and neutron-proton (panel (b))
cross-sections, calculated in accordance with
Eq.~(\ref{eq:cross_tot}). Line styles are the same as in
Fig.~\ref{F:sigmadiffstates}. Both neutron-neutron and
neutron-proton cross-sections are suppressed by in-medium effects
at smaller values of $E_{\rm c.m.}$ (or, equivalently, c.m.
momentum $p$) and become higher than the free-space cross-sections
at higher values of energy. Note that the neutron-proton total
cross-section in the Fig.~\ref{F:sigmatotstates}~(b) have sense
only for the symmetric nuclear matter, remind the discussion in
Sec.~\ref{S:crossect}.

\subsection{Transport coefficients}

\begin{figure}
\includegraphics[width=0.8\columnwidth]{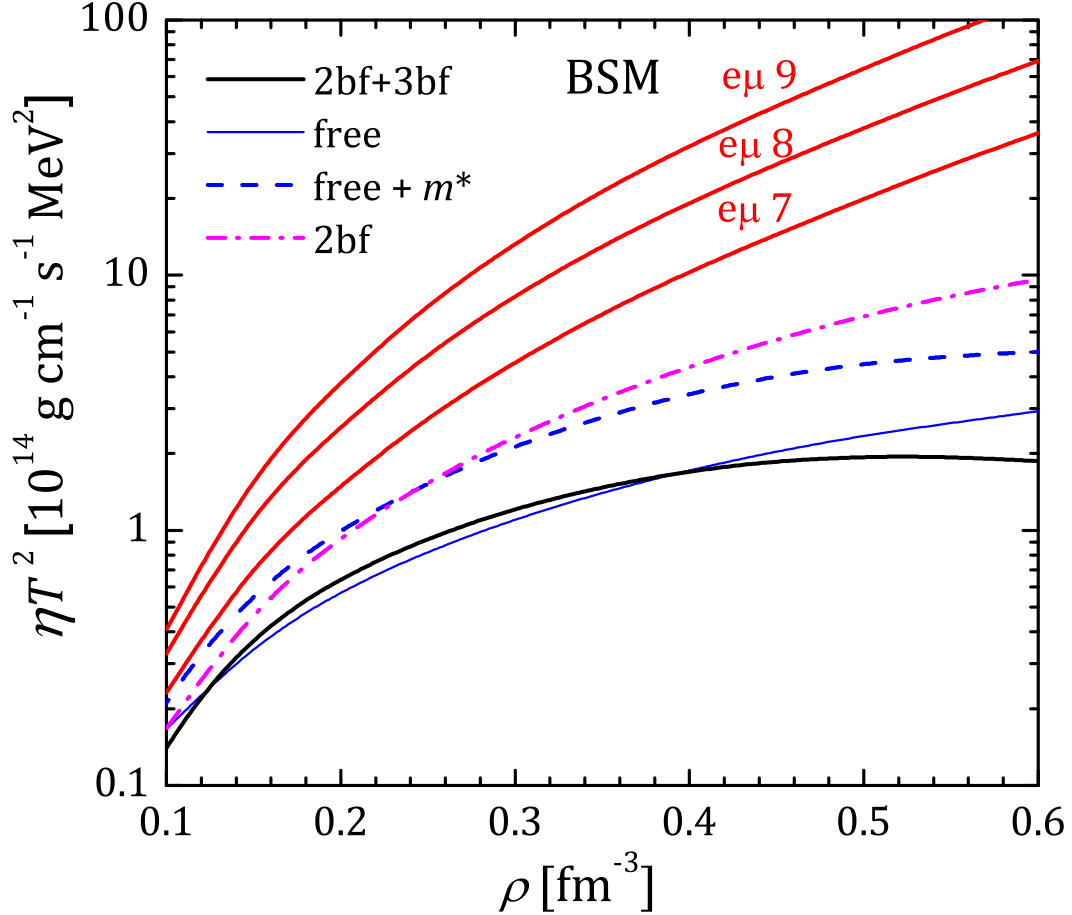}
  \caption{(color online) Shear viscosity of neutrons versus density in beta-stable nuclear matter.
  Solid line is calculated including all effects discussed in the text. Dash-dotted line represent contribution of two-body forces alone.
  Thin solid line shows the free-space result, while dashed line is for free space matrix elements, with inclusion of in-medium effective masses.
  Lines marked e$\mu$ show electron and muon contribution for three values of temperature $T=10^7$, $10^8$, and $10^9$~K (logarithm of temperature
  is noted near the curves).
  }\label{F:etaBSM}
\end{figure}

\begin{figure}
\includegraphics[width=0.8\columnwidth]{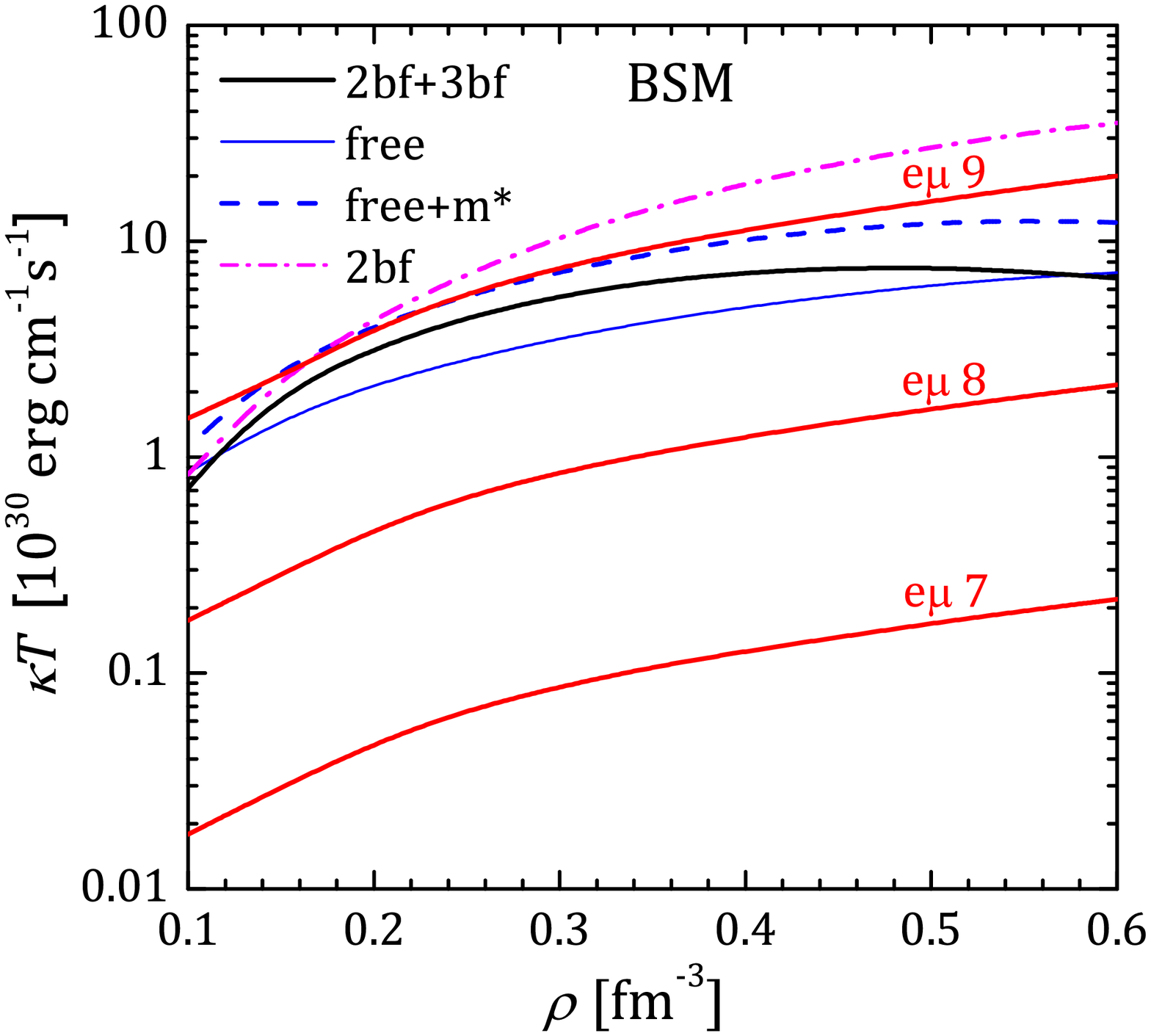}
  \caption{(color online) Thermal conductivity of neutrons versus density
  in beta-stable nuclear matter. Notations are the same as in Fig.~\ref{F:etaBSM}.}\label{F:kappaBSM}
\end{figure}

Figures \ref{F:etaBSM} and \ref{F:kappaBSM} show the neutron shear
viscosity and thermal conductivity, respectively, in the
beta-stable nuclear matter. Both quantities are given by
temperature-independent combinations $\eta T^2$ and $\kappa T$.
The results of full calculations, including two-body and
three-body forces are shown by solid lines. Note that here the
exact solutions of corresponding systems of kinetic equations are
presented. The results obtained neglecting three-body contribution
are given by dash-dotted lines in Figures~\ref{F:etaBSM} and
\ref{F:kappaBSM}, and the results of free-space calculations are
given by thin solid lines. Behavior of both shear viscosity and
thermal conductivity is similar. We find that in-medium effects on
two-body level increase the kinetic coefficients, while inclusion
of UIX three-body force works in opposite direction, decreasing
the values back close to the free-space results. To track the
effect of effective mass we also plot in Figs.~\ref{F:etaBSM} and
~\ref{F:kappaBSM} results of calculations with in-medium effective
masses, but free-space scattering matrix elements. These curves go
considerably higher than the results of full calculations.
Therefore, the in-medium effects on the scattering matrix are as
important as the effects of effective mass. In addition, in
Figs.~\ref{F:etaBSM} and \ref{F:kappaBSM} we plot the electron and
muon shear viscosity $\eta_{e\mu}$ and $\kappa_{e\mu}$,
respectively, in accordance with Refs.~\cite{sy07,sy08b}. These
quantities has non-standard temperature dependence
($\eta_{e\mu}\propto T^{-5/3}$, $\kappa_{e\mu}\propto T^{-1}$ in
the leading order), therefore combinations $\eta_{e\mu}T^2$ and
$\kappa_{e\mu}T$ are no longer temperature independent
\cite{sy07,sy08b}. We consider three values of temperature
$T=10^7$, $10^8$, and $10^9$~K, where the $\log_{10} T$~[K] is
shown near the corresponding curves. We see that neutron shear
viscosity in our model is much smaller than electron and muon one
for all densities and temperatures of consideration. This is due
to suppression effect of three-body forces. \pss{Our} result
contradicts the results of other authors \cite{bv07,zlz10,bpvv10}.
This issue will be discussed separately below. For the thermal
conductivity, situation is opposite. The relation $\kappa_{
n}\gg\kappa_{e\mu}$ is valid for all temperatures, except
\pss{for} the highest $T\sim 10^9$~K where neutron and
electron-muon contributions become comparable.
\begin{figure}
\includegraphics[width=0.8\columnwidth]{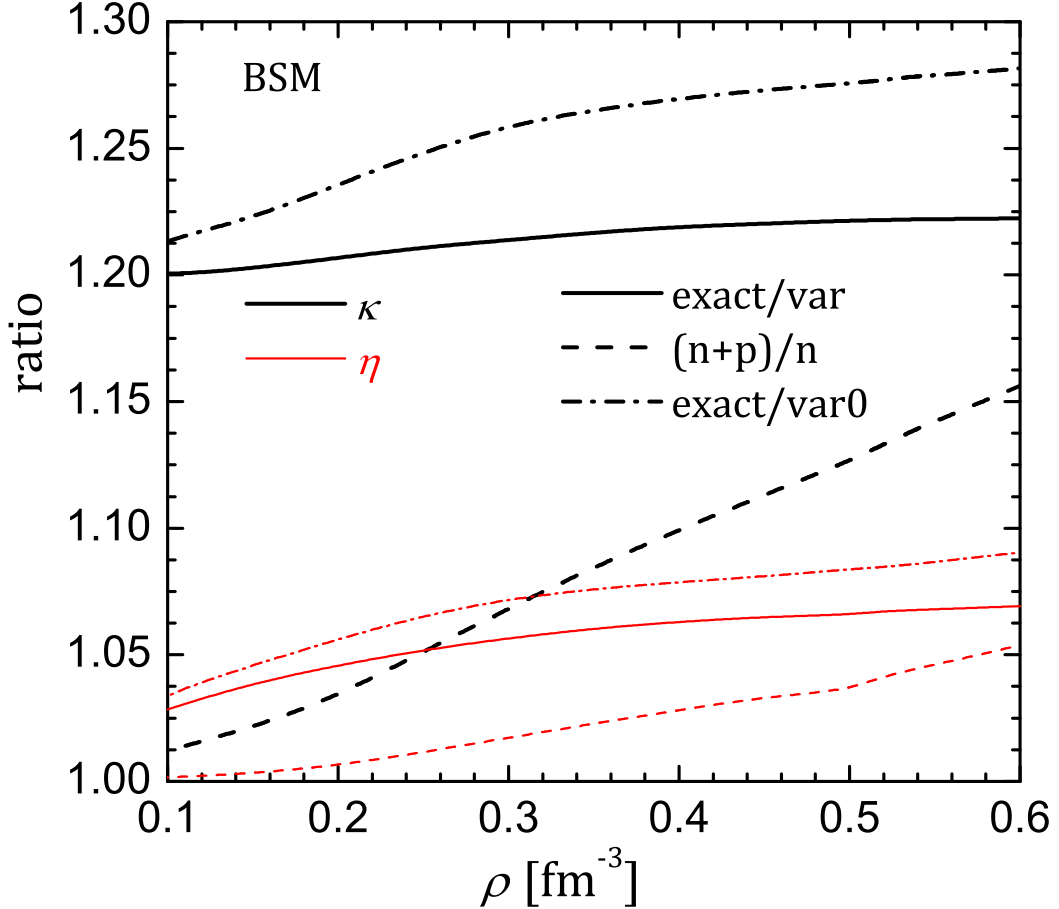}
  \caption{(color online) Ratio between exact and variational
  values of the kinetic coefficients (solid lines). Thick lines are for thermal conductivity, while thin
  lines are for shear viscosity. Dashed lines show the relative
  contribution of protons. Dash-dotted lines give the ratio
  between the exact values and those obtained in variational
  approximation with only neutrons as carriers.
  }\label{F:exact}
\end{figure}

In Fig.~\ref{F:etaBSM} and \ref{F:kappaBSM} the results of the
exact solution of the $2\times 2$ system of kinetic equations for
neutron-proton subsystem are presented. It is instructive to
compare the exact solution to simpler variational calculations
which solve simpler algebraic system (\ref{eq:var_kin}). This is
done in Fig.~\ref{F:exact} where the ratio between exact and
variational solutions is given by thick solid line for $\kappa$
and thin solid line for $\eta$, respectively. We see, that it is
enough to employ simple variational expressions with correction
factor $C_\kappa=1.2$ and $C_\eta=1.05$. In the same figure with
dash-dotted lines we compare the exact result with variational
approximation in which protons are only considered as scatterers.
In this case $2\times 2$ system (\ref{eq:var_kin}) reduces to one
equation for the neutron effective relaxation time. In this case
the correction coefficients are higher. Finally, we investigate
the proton contribution to the kinetic coefficients by plotting
the ratios $(\eta_{n}+\eta_{ p})/\eta_{n}$ and $(\kappa_{
n}+\kappa_{ p})/\kappa_{ n}$ with thin and thick dashed lines,
respectively. It is clear that the proton contribution to shear
viscosity can always be neglected. For the thermal conductivity
the proton contribution can reach 15\% at highest considered
density and can be included in the calculations.

\begin{figure}
\includegraphics[width=0.8\columnwidth]{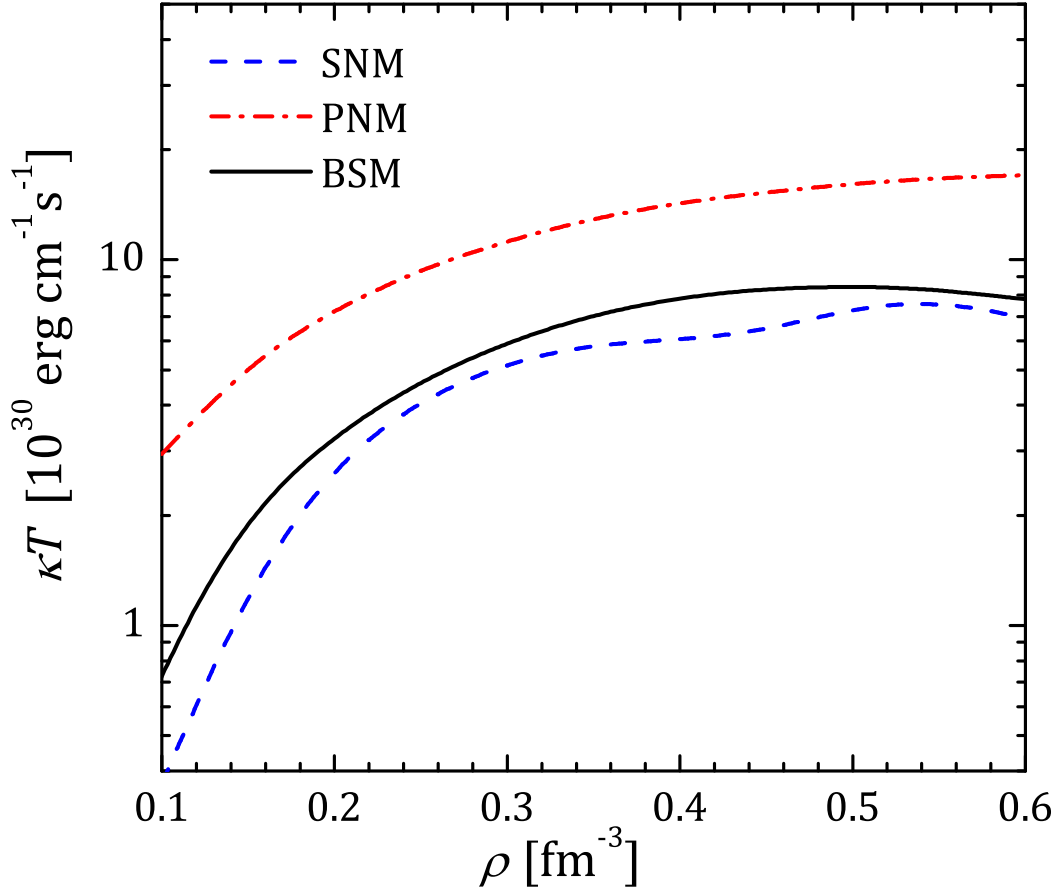}
  \caption{(color online)
  Total thermal conductivity ($\kappa_{ n}+\kappa_{ p}$) as a function of
  density for pure neutron matter (dash-dotted curves), symmetric
  nucleon matter (dashed curves), and beta-stable nucleon matter
  (solid curves).
  }\label{F:kappaDM}
\end{figure}
\begin{figure}
\includegraphics[width=0.8\columnwidth]{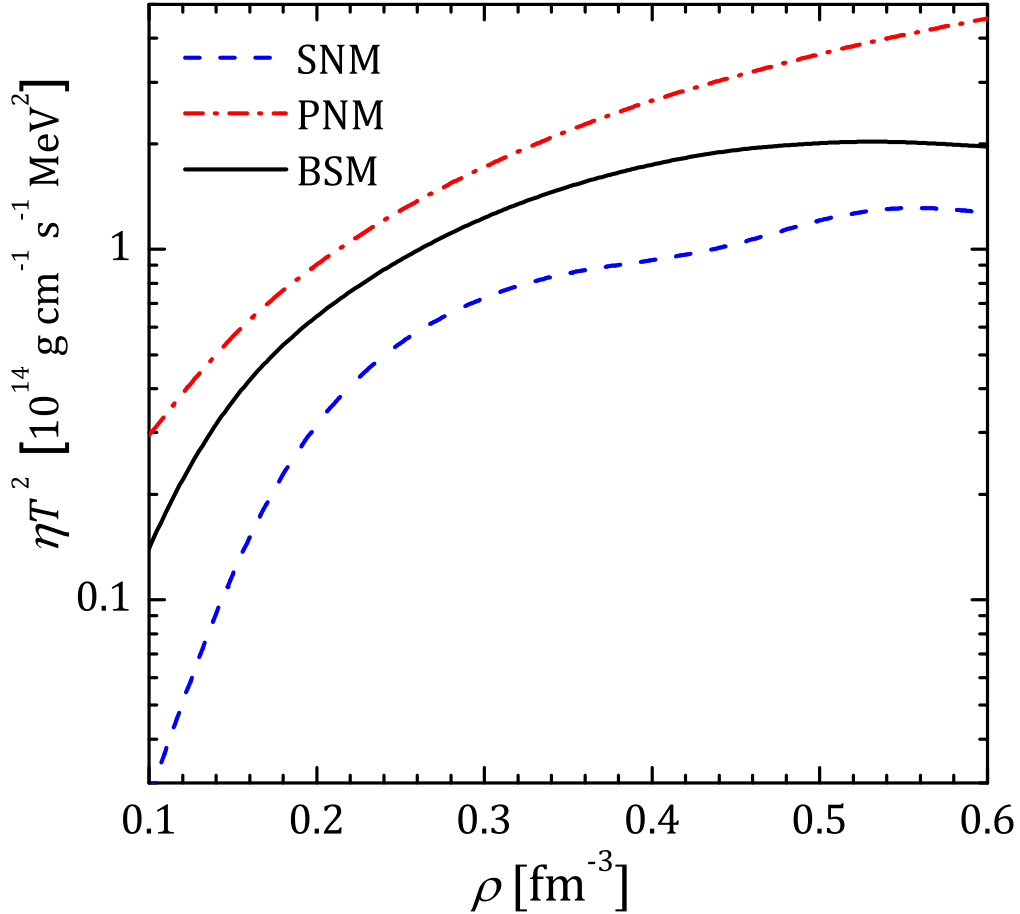}
  \caption{(color online)
  Total shear viscosity ($\eta_{n}+\eta_{p}$) as a function of
  density for pure neutron matter (dash-dotted curves), symmetric
  nucleon matter (dashed curves), and beta-stable nucleon matter
  (solid curves).
  }\label{F:etaDM}
\end{figure}
Finally, in Figs.~\ref{F:kappaDM} and \ref{F:etaDM} total  thermal
conductivity $\kappa_{n}+\kappa_{ p}$ and shear viscosity
$\eta_{n}+\eta_{ p}$, respectively, are plotted for three states
of matter -- pure neutron matter, symmetric neutron matter, and
beta-stable neutron matter. We see that even a small amount of
protons (as happens in beta-stable matter, see Fig.~\ref{F:xp})
leads to considerable reduction of kinetic coefficients. This
effect is more pronounced for the thermal conductivity
(Fig.~\ref{F:kappaDM}). It means that the neutron-neutron
collision frequencies and neutron-proton collision frequencies are
comparable despite low proton fraction. The reason for that lies
in the different kinematical restrictions for neutron-neutron and
neutron-proton collisions, as well as in the different behavior of
neutron-neutron and neutron-proton cross-sections  \cite{bhy01}.
Indeed, at small fraction of protons $p_{\rm Fp}\ll p_{\rm Fn}$
and neutron-proton scattering occurs at small c.m. angles
($q<q_m(P\approx p_{\rm F n})\ll p$, see Eq.~(\ref{eq:Skappa_ci})
and Eq.~(\ref{eq:Seta_ci})). In this case forward-scattering part
of the np cross-section plays the major role. In contrast, the
neutron-neutron collisions occur in whole
 range of c.m. angles ($q<q_m(P)=2 p$ in
Eqs.~(\ref{eq:Skappa_cc}) and (\ref{eq:Seta_cc})). In this case
effective collision frequencies are determined mainly by the nn
cross-section at large angles. \pss{Let us remind}, that due to
inclusion of $T_z=0$ isospin channel, \pss{the} np cross-section
is larger than \pss{the} nn \pss{one}. In addition, np
cross-section at small scattering angles is considerably increased
in comparison to that at large angles. Finally, smaller values of
energy give main contribution to \pss{the} np scattering in
comparison with energies \pss{relevant for the} nn scattering,
which additionally increases the contribution from \pss{the} np
scattering.
\begin{figure}
\includegraphics[width=0.8\columnwidth]{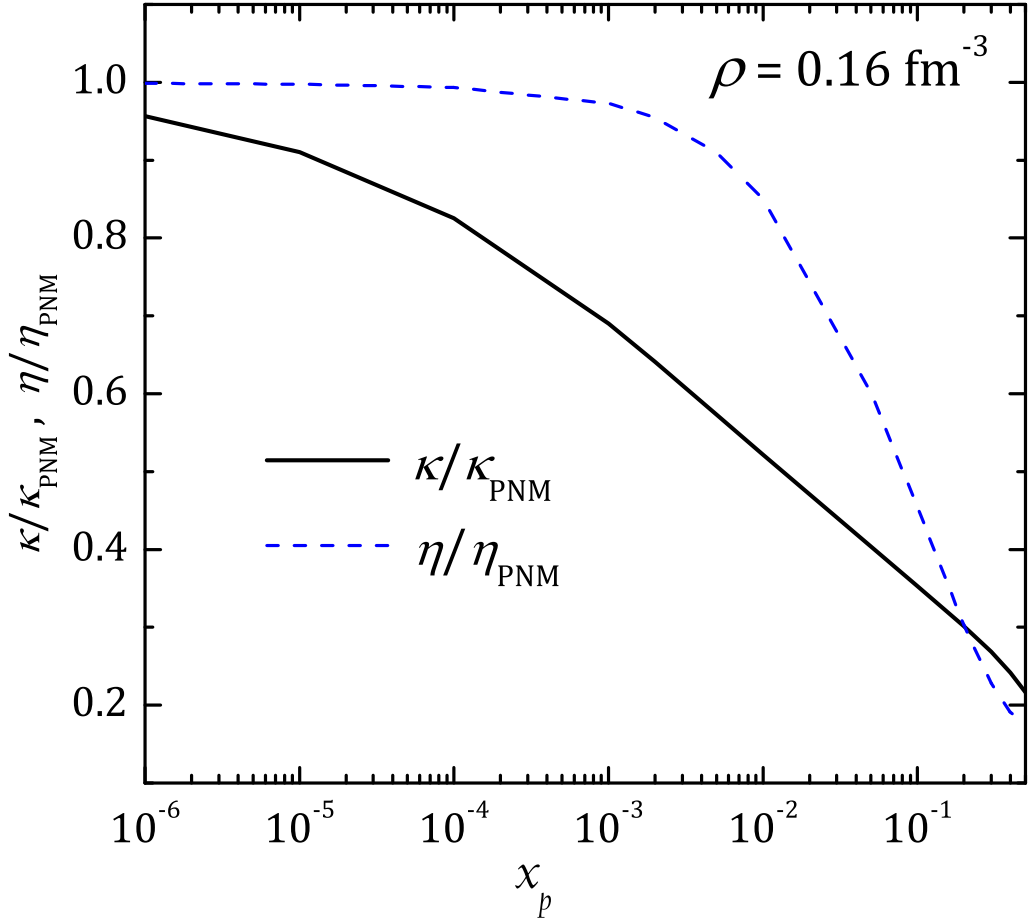}
  \caption{(color online) Ratios $\kappa/\kappa_{\rm PNM}$ (solid lines) and
  $\eta/\eta_{\rm PNM}$ (dashed lines) of kinetic coefficients in
   nucleon matter to corresponding quantities in pure neutron
   matter as a function of proton fraction $x_p$ at density $\rho=0.16$~fm$^{-3}$. Free-space
   scattering probabilities and bare masses are used.
  }\label{F:kin_xp}
\end{figure}
We illustrate the importance of \pss{the} neutron-proton
scattering by plotting in Fig.~\ref{F:kin_xp} ratios
$\kappa/\kappa_{\rm PNM}$ and $\eta/\eta_{\rm PNM}$ of total
neutron and proton thermal conductivity and shear viscosity to the
same quantities, calculated for the pure neutron matter, as a
function of the proton fraction $x_{ p}$ at the baryon density
$\rho=0.16$~fm$^{-3}$. For simplicity, in Fig.~\ref{F:kin_xp} we
used in-vacuum scattering probabilities and effective masses. One
can indeed observe, that for  very small proton fractions
neutron-proton scattering still contribute significantly to the
thermal conductivity. For the shear viscosity this effect is
weakened due to additional factor $q^2$ in expressions for
effective collision frequencies (compare Eq.~(\ref{eq:Seta_ci})
and Eq.~(\ref{eq:Skappa_ci})). In both cases, at $x_{p}>0.1$ the
use of kinetic coefficients of pure neutron matter as an estimate
 of true values can lead to an error \pss{by a} factor more than 2.

\subsection{Comparison with results of other authors}\label{S:compar}

Let us compare our results with the most recent calculations by
Zhang et al. \cite{zlz10} and Benhar et al.
\cite{bv07,bpvv10,cb11}.
\begin{figure}[t]
\includegraphics[width=0.8\columnwidth]{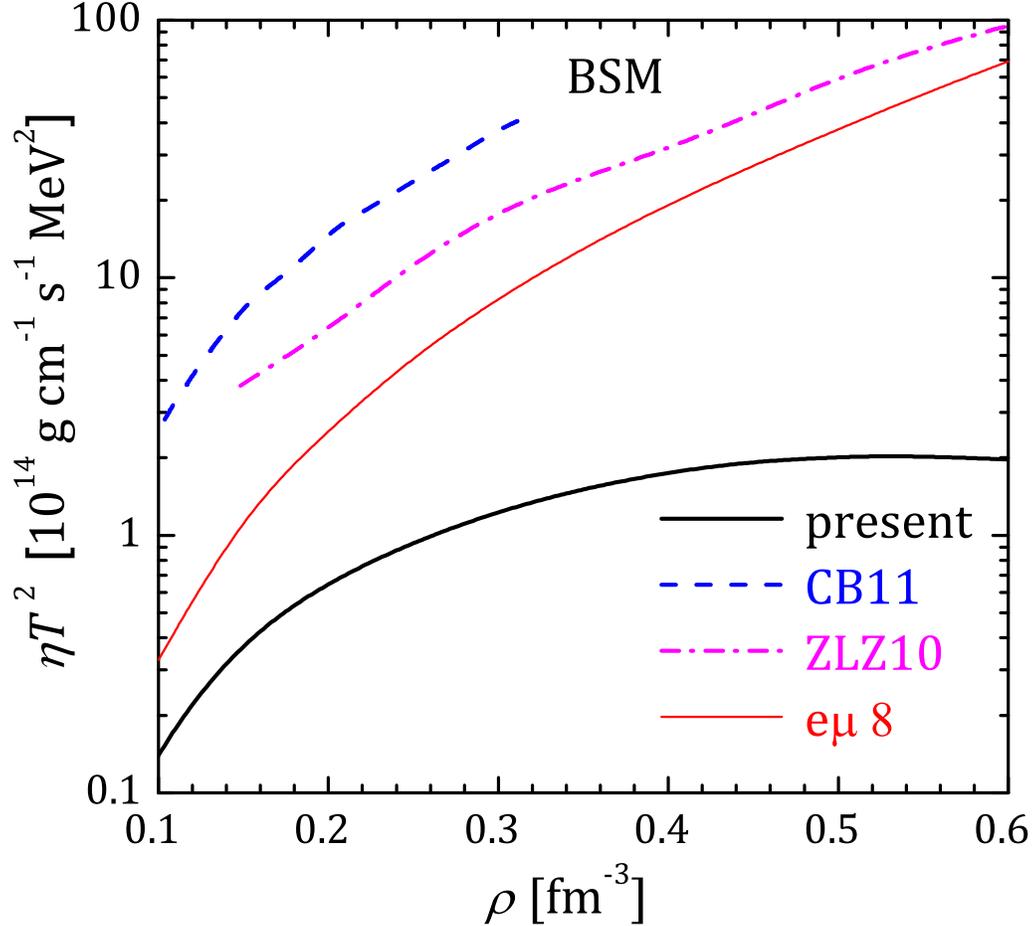}
  \caption{(color online) Comparison of the shear viscosity in
  beta-stable nuclear matter obtained by different groups. Solid
  line -- present calculations, dashed line -- calculations by
  Carbone and Benhar \cite{cb11}, dash-dotted line -- those by
  Zhang et al. \cite{zlz10}. Thin solid line shows electron and
  muon shear viscosity at $T=10^8$~K.
  }\label{F:eta_beta_comp}
\end{figure}
Figure~\ref{F:eta_beta_comp} shows the shear viscosity $\eta T^2$
for the beta-stable nuclear matter as presented by these groups in
comparison with the present work (solid line). The results from
Ref.~\cite{cb11} are given by dashed line and results from
Ref.~\cite{zlz10} by dash-dotted line. For comparison $\eta_{\rm
e\mu}T^2$ is shown for $T=10^8$~K. Apart from the fact that all
authors use different equation of state and therefore different
proton fractions, and the methods of calculation are different the
results in Fig.~\ref{F:eta_beta_comp} disagree with each other.
The similar situation is observed in case of the thermal
conductivity as shown in Fig.~\ref{F:kappa_beta_comp}.
\begin{figure}
\includegraphics[width=0.8\columnwidth]{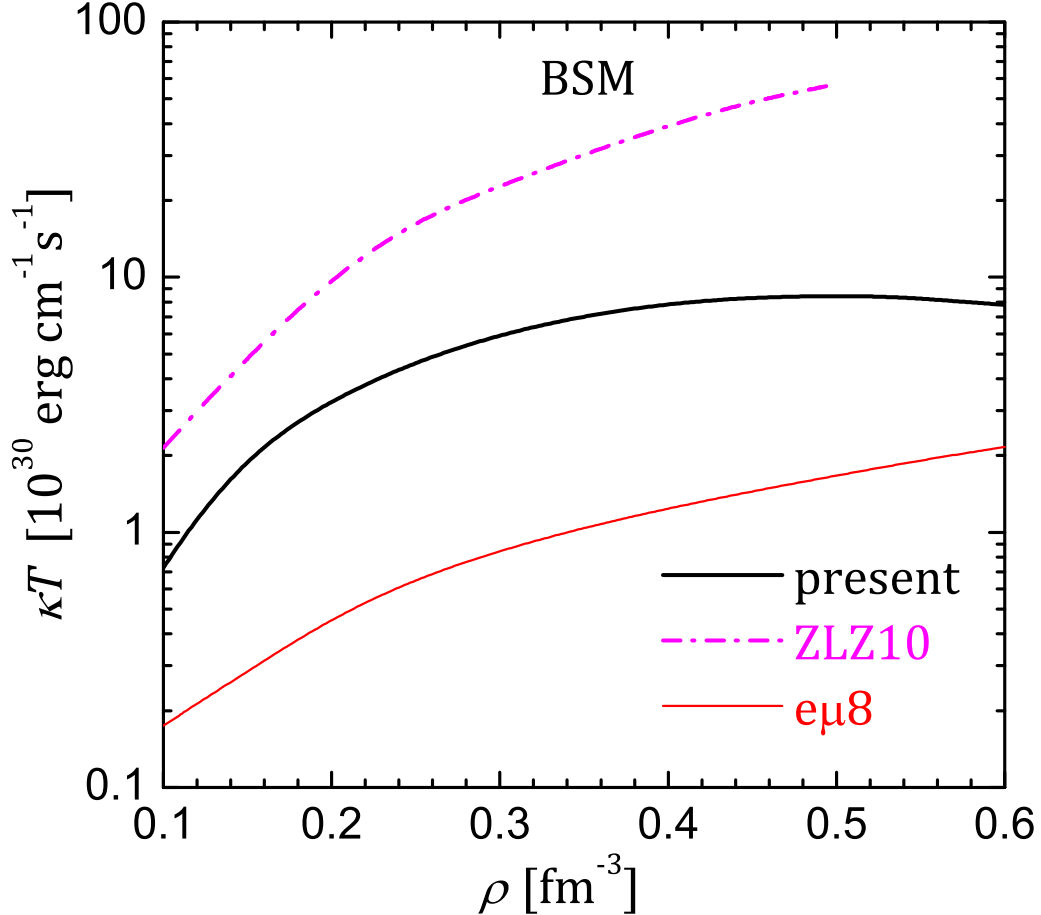}
  \caption{(color online) Comparison of the thermal conductivity in
  beta-stable nuclear matter obtained by different groups.
  Notations are same as in Figure~\ref{F:eta_beta_comp}.
  }\label{F:kappa_beta_comp}
\end{figure}
Unfortunately, the calculations of Benhar et al. for the thermal
conductivity in beta-stable nuclear matter are not available.
However,  Zhang et al. \cite{zlz10} results (dot-dashed curve in
Fig.~\ref{F:kappa_beta_comp}) lie much higher than the present
calculations.

\begin{figure}[th]
\includegraphics[width=0.8\columnwidth]{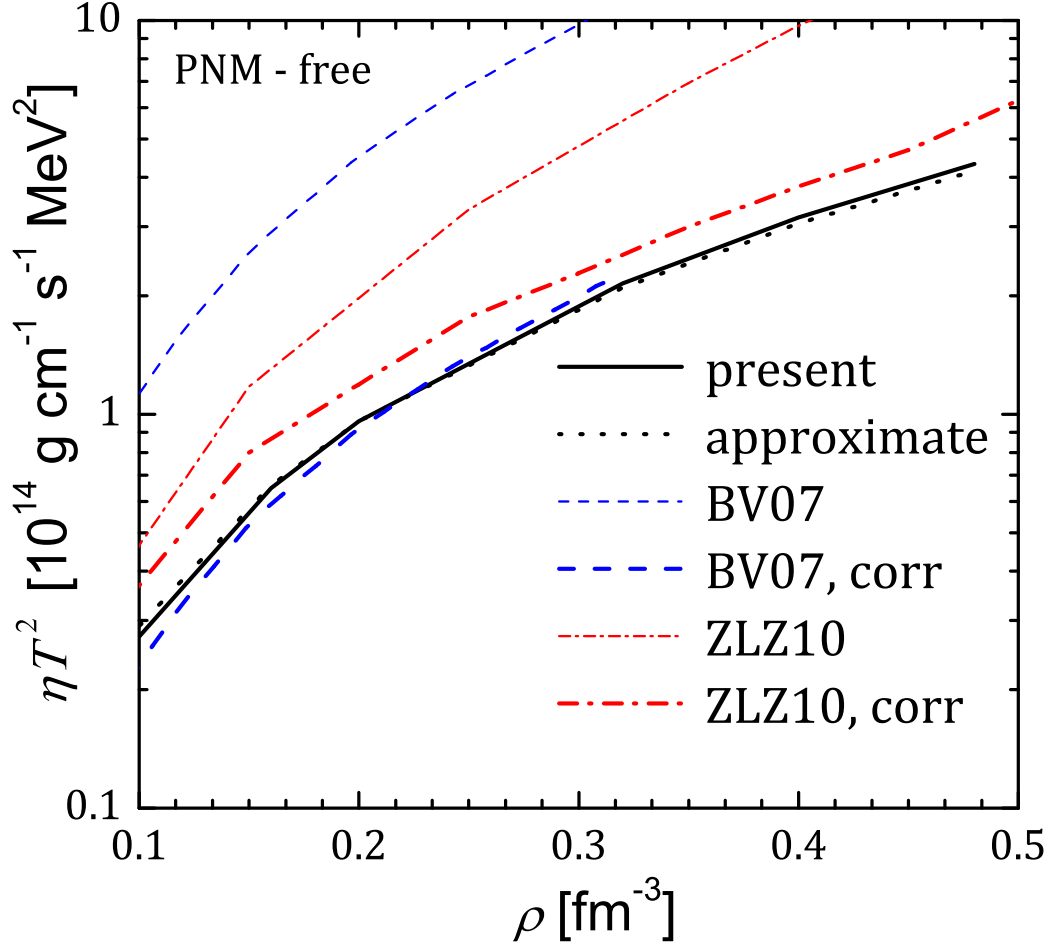}
  \caption{(color online) Comparison of the shear viscosity in PNM
  in free scattering approximation as reported by different groups.
  Solid line shows the results of the present calculations, dotted
  line correspond to the approximate expression. Thin dashed and
  dash-dotted lines show the results of Ref.~\cite{bv07} and
  Ref.~\cite{zlz10}, respectively. Thick lines of the corresponding type show the
  corrected results, see text for details.
  }\label{F:eta_free_comp}
\end{figure}

\begin{figure}[bh]
\includegraphics[width=0.8\columnwidth]{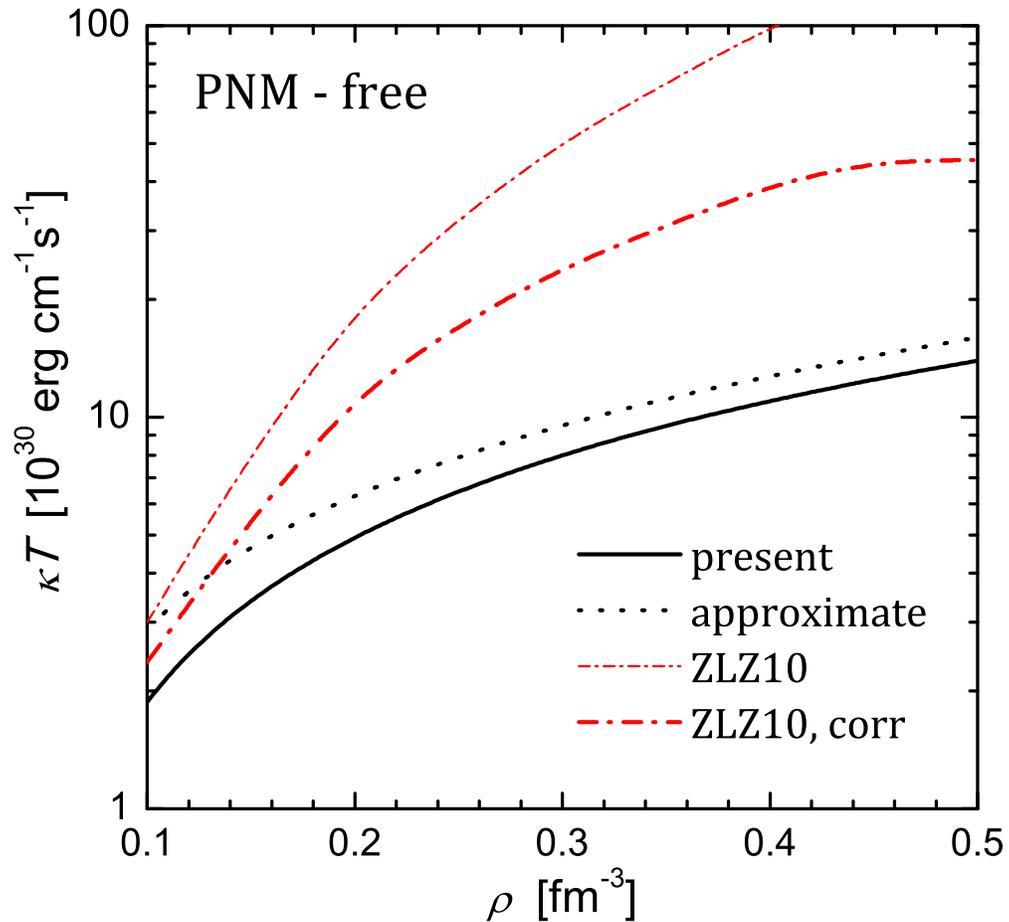}
  \caption{(color online) Comparison of the thermal conductivity
  in PNM in
  free scattering approximation as reported by different groups.
 Notations are the same as in Fig.~\ref{F:eta_free_comp}.
  }\label{F:kappa_free_comp}
\end{figure}

In attempt to find the source of this huge discrepancy we analyzed
in detail the models used in the series of papers
\cite{bv07,bpvv10,cb11} and \cite{zlz10,zang07}. To begin with we
consider \pss{the} simpler case of pure neutron  matter. In this
case it is possible to construct the approximate expressions for
the shear viscosity and thermal conductivity  which \pss{depend
only on the total neutron-neutron cross-section at the c.m. energy
on the Fermi surface $E_{\rm c.m.}=p_{\rm Fn}^2/m_{ N}$}, see
Appendix~\ref{A:kinapprox} for details. It is important, that
these approximate expressions provide an independent test of the
calculations for the free-scattering case, where the in-vacuum
cross-section is used. As the total \pss{in-vacuum} nn
cross-section is known relatively well, the ``free'' result of any
calculation must lie close to the values obtained from the
approximate expressions. The results of our calculations, as well
as those in Refs.~\cite{bh99,bhy01}, satisfy this criterion. Zhang
et al. \cite{zlz10} present both shear viscosity and thermal
conductivity in PNM free case, while Benhar et al. \cite{bv07,
bpvv10} show only the shear viscosity for the free case. The
results of both groups, at the first glance, strongly disagree
with the results presented here, and do not agree with the
approximate expression (see Fig.~\ref{F:eta_free_comp} and
\ref{F:kappa_free_comp}). The careful examination of the
expressions in Refs.~\cite{zlz10,bv07,bpvv10} shows that in the
free case the authors still incorporate some in-medium effects
through the Fermi velocities in the definitions of the kinetic
coefficients in eqs.~(2) and (3) in Ref.~\cite{zlz10}, eq.~(1) in
Ref.~\cite{bv07}, eq.~(24) in Ref.~\cite{bpvv10}. The factor
$v_F^2$ leads to the appearance of squared effective mass in the
expressions for the kinetic coefficients. In addition,
eqs.~(2)--(6) of Benhar and Valli \cite{bv07} \pss{indicate} that
these authors lost the factor $\pi$ in the expression for the
averaged scattering probability, remind the remark in the end of
the Sec.~\ref{S:kincoeff} and Eq.~(\ref{eq:W_cross}). Therefore
their result and the results of the subsequent papers
\cite{bpvv10,cb11} are overestimated by the factor of $\pi$.
Therefore we corrected the results of Ref.~\cite{zlz10} by the
factor $m_{n}^{*2}$, where neutron effective mass is taken from
the fig.~2 in \cite{zlz10} and the result of \cite{bv07} for
viscosity by the factor $m_{ n}^{*2}/\pi$. The latter effective
mass is not available from the references and we assumed $m_{
n}^*\approx 0.8 m_{ N}$. This value is consistent with the
dot-dashed curve in fig~1(a) in Ref.~\cite{bv07} which is reported
to be obtained from eqs.~(43) and (46) of Ref.~\cite{bh99} with
the same effective mass as used elsewhere. Note, that expressions
in Ref.~\cite{bh99} contain the fourth power of $m_{n}^*$ as those
authors used the in-vacuum transition probability, while authors
of Ref.~\cite{bv07} use the in-vacuum cross-section (their
eq.~(6)). The discrepancy between the results of Benhar and Valli
\cite{bv07} and Baiko and Haensel \cite{bh99} for the ``free
scattering'' approximations is due to the different effective mass
power and the $\pi$ factor, not due to the correction factor to
the variational solution as were incorrectly assumed by authors of
Ref.~\cite{bv07}. In fact, this correction, although small for the
shear viscosity, is included in eqs.~(43) and (46) in
Ref.~\cite{bh99}.


The ``corrected'' values of  shear viscosities are shown in
Fig.~\ref{F:eta_free_comp} with thick dashed line for Benhar and
Valli calculations and with thick dash-dotted line for Zhang et
al. calculations. The uncorrected values are shown with thin
lines. The results of the present paper are shown with the solid
line, while the approximate result is shown by the dotted line and
nearly \pss{coincides} with the exact result. Note that we
multiplied the approximate variational expression by the
correction factor 1.05. One observes, that the results of Benhar
and Valli agree well with the approximate expression, while Zhang
et al. curve has qualitatively the same behavior, but gives
somewhat higher values. The different situation is observed for
the thermal conductivity, as shown in
Fig.~\ref{F:kappa_free_comp}. Again by the solid and dotted lines
we show the result of the present and approximate calculations.
Now the approximate expression (which is corrected by the factor
1.2) is less accurate than in \pss{the} case of the shear
viscosity. Nevertheless the results of Zhang et al. \cite{zlz10},
corrected by the $m_{n}^{*2}$ factor, still go much higher. In
order to check if this can be the result of the different model
used, let us look at the cross-sections reported in
Refs.~\cite{zlz10,zang07}. In Ref.~\cite{zang07} the Argonne v14
potential was used, while in Ref.~\cite{zlz10} authors used Bonn-B
potential. The reported cross-sections are close, as expected.
However, one can note that the nn cross-sections (for the free
scattering) in Ref.~\cite{zlz10,zang07} are different from those
obtained in the present paper. One difference is that these
authors clearly plot 1/2 of the differential $nn$ cross-sections,
so that the c.m. solid angle would be 4$\pi$. Another difference
is that the cross-sections have much stronger dependence on
$\theta_{\rm c.m.}$ at forward and backward scattering than those
obtained here, or in other works (for instance,
\cite{LM94,ffs01}). Although the reason for this difference is
unknown to us, we found that we can reproduce well the free-space
nn cross-sections in Refs.~\cite{zlz10} and \cite{zang07} (for the
corresponding potentials) if we omit the phase factor
$i^{\ell'-\ell+\bar{\ell}-\bar{\ell}'}$ in Eq.~(\ref{eq:Q_L}).
This factor comes from the partial wave expansion of the plane
waves. This affects only non-diagonal elements, therefore the
total cross-section does not change if this phase factor is
omitted. Hence the resulting kinetic coefficients still should be
close to the approximate expressions. We checked this explicitly,
using the cross-sections calculated without the phase factor. This
test allows to check the results of \cite{zlz10} irrespectively if
the phase factor omission is a real reason for such form of the
cross-section, or \pss{where} it is a coincidence. We found that
the both shear viscosity and thermal conductivity are still in
agreement with the corresponding approximate expressions, and did
not found anything similar to the plot in
Fig.~\ref{F:kappa_free_comp}. Therefore we conclude, that the
thermal conductivity calculations in Ref.~\cite{zlz10} are
incorrect and probably give overestimated values for this kinetic
coefficient, while the shear viscosity calculations of
Ref.~\cite{zlz10} and Refs.~\cite{bv07,bpvv10} (if divided by
$\pi$) look plausible.

Now let us return to the shear viscosity in
Fig.~\ref{F:eta_beta_comp}. Results of the in-medium calculations
by Zhang et al.~\cite{zlz10} and Carbone and Benhar~\cite{cb11},
even corrected by $\pi$, are much higher than our calculations.
These groups used different approaches, and we discuss them
separately.

Zhang et al.~\cite{zlz10} reports approximately order of magnitude
increase of the shear viscosity due to the medium effects. The
above analysis of their in-vacuum results
(Fig.~\ref{F:eta_free_comp}) suggests that in both ``free'' and
``in-medium'' calculations same Fermi velocities were used.
Therefore the increase in the in-medium viscosity is solely due to
the decrease in the in-medium cross-section. However, as follows
from the plots in Refs.~\cite{zlz10} and \cite{zang07} this
decrease is lower, at most by the factor of 2-3 in the high-energy
region of interest. Hence the increase in the shear viscosity if
calculated properly should be of the same order. \referee{Note
that for shear viscosity in PNM and SNM as well as for thermal
conductivity in all three states this increase is indeed reported
to be 2-4 in Ref.~\cite{zlz10}.} Therefore we believe that the
in-medium shear viscosity \referee{for BSM} in Ref.~\cite{zlz10}
is calculated inaccurately and is overestimated by a factor of
3-5. The remaining difference between our shear viscosity and one
by Zhang et al.~\cite{zlz10} would be still about a factor of 3-5,
but the latter difference is due to the different physical model
used. Indeed, as already stated before, they used different model
for three-body force, which \pss{led} to the decrease of in-medium
cross-section, while the three-body force we employ operates in
opposite directions. The second source of difference is the
rearrangement modifications of the effective masses. The inclusion
of the rearrangement led Zhang et al.~\cite{zlz10,zang07} to a
strong reduction of the effective masses, \pss{while} in contrast
we observe the increase of effective masses due to three-body
forces (see Fig.~\ref{F:meff}). These factors again work in the
opposite \pss{directions}. Nevertheless, the use of the modified
effective mass is cautioning. Indeed, the origin of the
rearrangement contribution lies in the functional dependence of
the $G$-matrix on the occupation numbers. This leads to difference
between Landau quasiparticle energy $\epsilon_{ L}(p)$ given by
the functional derivative of the total energy of the system with
respect to the distribution function and the Brueckner
self-consistent single-particle potential $\epsilon(p)$ in
Eq.~(\ref{eq:spp}). The two quantities would coincide provided
that the functional derivative acting on the $G$-matrix is
neglected. Therefore the quasiparticle effective mass which is
found from $\epsilon_{ L}(p)$ deviates from the Brueckner
effective mass $m^*$. Physically, including rearrangement
corrections incorporates to some extent the effects of the medium
polarization. However, for consistency, the same effects should be
included in the quasiparticle interaction. Therefore this
interaction (and quasiparticle scattering amplitude as well) must
differ from the Brueckner $G$-matrix at the same footing. Provided
strong modification of the effective mass one could expect strong
modification of the quasiparticle scattering amplitude at the same
level of approximation. The size and the direction (increase or
decrease) of this effect is unknown and requires a separate
consideration, which lies outside the scope of the present paper.
It is possible that it can suppress the kinetic coefficients more,
or, in opposite, there is a possibility of counter-compensation of
the effective mass effect.

In a series of papers \cite{bv07,bpvv10,cb11} the correlated basis
function (CBF) formalism was used which is different from the BHF
approach. However, in Ref.~\cite{bpvv10} authors compared results
obtained in CBF and $G$-matrix approaches and obtained overall
agreement. The reasons of the huge in-medium increase of kinetic
coefficients in these works are different than in Zhang et
al.~\cite{zlz10}. First of all, we note that the three-body force
effects are reported to be small. In fact, the comparison of the
results in fig.~1.(b) in Ref.~\cite{bv07} and in fig.~5 in
Ref.~\cite{bpvv10} where the three-body forces, as written, are
not included, shows that the three-body forces even slightly
decrease the shear viscosity. Therefore the strong in-medium
increase of the shear viscosity (again $\sim 10$) is due to the
squared effective mass, which gives a factor of 2, and due to
decrease of the cross-section, which is a factor of 5, according
to fig. 4 in Ref.~\cite{bpvv10} (note that the free-space
cross-section reported in the same figure is underestimated by the
factor $\sim$1.5 with respect to the true values). Let us note,
that we do not observe so strong decrease of the in-medium
cross-section at the two-body level (Fig.~\ref{F:sigmatotstates}).
Same smaller decrease (factor of 2) is also found by the other
authors (e.g. \cite{LM94, ffs01}), including Zhang et al.
\cite{zang07}. In addition, the proton fraction used by Carbone
and Benhar~\cite{cb11} is slightly lower than one we use.

Finally, we stress that inaccuracies in the recent calculations of
the two groups do not allow one to deduce which percentage of the
effect is due to the selection of the physical model and which
part is related to these inaccuracies.

\section{Conclusions}
\label{S:conclusion}

We have calculated thermal conductivity and shear viscosity of
nuclear matter in beta-equilibrium. The neutron and proton
interaction was described by the Argonne v18 potential with
inclusion of effective Urbana IX three-body forces. The scattering
of particles were treated in the non-relativistic Brueckner-Hartree-Fock
approximation with the continuous choice of the single-particle
potential.

Our main results are as follows
\begin{enumerate}
\item Shear viscosity and thermal conductivity of nuclear matter
are modified by the medium effects in comparison with the values
obtained from the free-space cross-sections. However this
modification is not so strong as reported previously.
\item The medium effects of the renormalization of the squared
matrix element due to the Pauli blocking in the intermediate
states and of the effective mass via the reduction of the density
of states are comparable and cannot be separated.
\item The Urbana IX three-body forces lead to \pss{the} increase
of the scattering probabilities and, therefore, to \pss{a}
considerable reduction of the kinetic coefficients in comparison
to the two-body case. The effect of the three-body forces is
sizable at $\rho\gtrsim0.15$~fm$^{-3}$.
\end{enumerate}

The question remains how the kinetic coefficients would depend on
the particular model for the three-body force. It is clear that
they are more sensitive to change of the model than the equation
of state. The results of Ref.~\cite{zlz10} suggest that the
inclusion of different three-body forces could lead to increasing
of the kinetic coefficients in contrast to the results of the
present paper. The investigation of the model-dependence of the
values of the kinetic coefficients is a good project for the
future.

In all our calculations we used the non-relativistic BHF
framework. The non-relativistic approach could be questioned,
especially at high density. However it has been shown
\cite{brownetal87} that the main relativistic effect, as included
in the Dirac-Bruckner (DBHF) scheme, is equivalent to the
introduction of a particular three-body force at the
non-relativistic level. Therefore the use of a TBF in BHF
calculations incorporates in an effective way the relativistic
corrections.

Finally let us note, that we have neglected the possible effects
of superfluidity. It is believed that the neutrons and protons in
the neutron star cores can be in the superfluid state. The
critical temperatures of the superfluid transition are very
model-dependent and can vary as $T_{c}\sim (10^8-10^{10})$~K (see,
for example, \cite{ls01}). The effects of superfluidity on the
kinetic coefficients were considered in the approximate way in
Refs.~\cite{bhy01, sy08b}. \pss{The investigation of these effects
 is outside the scope of the present paper.}

\begin{acknowledgments}
This work was partially supported by the Polish NCN grant no
2011/01/B/ST9/04838 and INFN project CT51. PSS acknowledge support
of the Dynasty Foundation, RFBR (grant 11-02-00253-a), RF
Presidential Programm NSh-4035.2012.2, and Ministry of Education
and Science of Russian Federation (agreement No.8409, 2012).
\end{acknowledgments}

\bibliography{../../BIBNS/NeutronStars}


\appendix

\section{Angular integrations}\label{A:angular}
We deal with the integrals of the form
\begin{equation}
  {\cal Q}^{(ij)} =
  \int\limits_{|p_{\rm Fc}-p_{\rm Fi}|}^{p_{\rm Fc}+p_{\rm Fi}} {\rm d} P
  \int\limits_0^{q_m(P)} {\cal Q}(P,q) \frac{P^i
  q^j\,{\rm d} q}{\sqrt{q_m^2-q^2}},
\end{equation}
where ${\cal Q}(P,q)$ is expanded in the Legendre polynomials
\begin{equation}
  {\cal Q}(P,q)=\sum_L {\cal Q}_L(P) {\cal
  P}_L\left(1-\frac{q^2}{2p^2}\right).
\end{equation}
Then the internal integral (over $q$) can be obtained analytically
\begin{widetext}
\begin{equation}
  \int\limits_0^{q_m} \frac{q^j {\rm d} q}{\sqrt{q_m^2-q^2}}{\cal
  P}_L\left(1-\frac{q^2}{2p^2}\right)=\frac{q_m^j}{2}B\left(\frac{j+1}{2},\frac{1}{2}\right)
  {}_3F_{2}\left(-L,L+1,\frac{j+1}{2};1,\frac{j}{2}+1;\frac{q_m^2}{4p^2}\right),
\end{equation}
\end{widetext}
where $B\left(\frac{j+1}{2},\frac{1}{2}\right)$ is the
beta-function, and ${}_3F_2$ is the generalized hypergeometric
function. The latter function, in fact, reduces to the $L-1$ order
polynomial in $q_m^2/(4p^2)$, as its first argument, $-L$, is a
negative integer.

\section{Approximate expressions for neutron kinetic
coefficients}\label{A:kinapprox}
 The simplest variational expressions for thermal conductivity
and shear viscosity of one-component Fermi-liquid read \cite{bp91}
\begin{eqnarray}
  \kappa_{\rm var}&=&\frac{20\pi^2 p_F^3}{9 m^{*4}\langle  W\rangle}\frac{1}{T}\left(
  3-\lambda_\kappa\right)^{-1},\\
  \eta_{\rm var}&=&\frac{4p_F^5}{5m^{*4} \langle  W\rangle}\frac{1}{(k_BT)^2}\left(1-\lambda_\eta\right)^{-1},
\end{eqnarray}
where $\lambda_\kappa$ and $\lambda_\eta$ encapsulate the
kinematical factors, namely
\begin{eqnarray}\label{eq:Wlambda}
  \langle W\rangle (1-\lambda_\kappa)=4\langle W \sin^2\frac{\theta}{2}
  \rangle,\\
  \langle W\rangle (1-\lambda_\eta)=3\langle W \sin^4\frac{\theta}{2} \sin^2
  \phi\rangle.\label{eq:Wlambda1}
\end{eqnarray}
The Abrikosov-Khalatnikov angle $\phi$ in the considered case is
equal to the c.m. angle $\theta_{\rm c.m.}$, and angle $\theta$ is
connected to the relative momentum as \referee{$p=p_F
\sin(\theta/2)$}. The scattering amplitude $W$ is connected to the
differential cross-section as
\begin{equation}\label{eq:W_cross}
  W=\pi {\cal Q}=\frac{16\pi^3\hbar^4}{m^{*2}} \frac{
  {\rm d}\sigma}{{\rm d}\Omega_{\rm c.m.}}.
\end{equation}
The Abrikosov-Khalatnikov averaging is defined as
\begin{equation}
  \langle W\rangle = \frac{1}{2\pi}\int_0^\pi {\rm d}\theta
  \sin(\theta/2) \int_0^{2\pi} {\rm d} \phi\, W(\theta,\phi).
\end{equation}
On can observe, that the main contribution to averages
(\ref{eq:Wlambda})--(\ref{eq:Wlambda1}) is given by the region of
$\theta\approx \pi$ due to powers of $\sin (\theta/2)$ in
kinematical factor. Therefore the result will be mainly determined
by the cross-section in the high-energy region $p\sim p_F$.
Assuming also that the angular structure of the cross-section is
flat (which is justified for neutron-neutron scattering at the
energies of interest) we can substitute
\begin{equation}
  \frac{{\rm d} \sigma}{{\rm d} \Omega_{\rm c.m.}} \to \frac{\sigma_{\rm
  tot}}{2\pi}.
\end{equation}
Note that for identical particles the possible scattering solid
angle is $2\pi$, not $4\pi$. Under these two approximation, the
final expressions are
\begin{eqnarray}
  \kappa_{\rm var}&\approx& \frac{5 p_F^3}{106 m^{*2} T}
  \left[\sigma_{\rm tot}(p=p_F)\right]^{-1} \label{eq:kappa_appr}\\
  \eta_{\rm var}&\approx& \frac{p_F^5}{16\pi^2 m^{*2} (k_B T)^2}\left[\sigma_{\rm
  tot}(p=p_F)\right]^{-1}.\label{eq:eta_appr}
\end{eqnarray}
For bare particles it is convenient to  write \pss{$E_{\rm
c.m.}=2p_F^2/m$} in the argument of the total cross-section
instead of $p=p_F$. Note that the approximations
(\ref{eq:kappa_appr})--(\ref{eq:eta_appr}) are good as long as
total cross section does not change significantly in the region of
large $\theta$. It is easy to write more general expression,
assuming only the flat angular dependence. In this case kinetic
coefficients would be determined by integration of the total
cross-section over the laboratory energy with corresponding
kinematical factors.

\end{document}